\documentclass[a4paper,fleqn,usenatbib]{mnras}
\usepackage{graphics}
\usepackage{graphicx}
\usepackage{soul}
\usepackage{multicol} 
\usepackage{amsmath}
\usepackage{amssymb}
\usepackage{url}
\usepackage{ulem}
\usepackage[T1]{fontenc}
\usepackage{ae,aecompl}
\usepackage{xcolor}
\usepackage{widetext}

\title[Imprint of massive neutrinos on PH of LSS] {Imprint of massive neutrinos on Persistent Homology of large-scale structure}
\author[M. H. Jalali Kanafi, S. Ansarifard and S. M. S. Movahed]
{M. H. Jalali Kanafi$^{1}$, S. Ansarifard$^{2}$ \& S. M. S. Movahed$^{1,3}$\thanks{E-mail: m.s.movahed@ipm.ir}
\\
$^{1}$Department of Physics, Shahid Beheshti University,  1983969411, Tehran, Iran\\
$^{2}$ School of Physics, Institute for Research in Fundamental Sciences (IPM), P. O. Box 19395-5531, Tehran, Iran\\
$^{3}$ School of Astronomy, Institute for Research in Fundamental Sciences (IPM), P. O. Box 19395-5531, Tehran, Iran
}

\begin{document}
    \maketitle

    \begin{abstract}
  
   Exploiting the Persistent Homology technique and its complementary representations, 
   we examine the footprint of summed neutrino mass ($M_{\nu}$) in the various density fields simulated by the publicly available Quijote suite. The evolution of topological features by utilizing the  super-level filtration on three-dimensional density fields at zero redshift, reveals a remarkable benchmark for constraining the cosmological parameters, particularly $M_{\nu}$ and $\sigma_8$. The abundance of independent closed surfaces (voids) compared to the connected components (clusters) and independent loops (filaments),  is more sensitive to the presence of $M_{\nu}$ for $R=5$ Mpc $h^{-1}$ irrespective of whether using the total matter density field ($m$) or CDM+baryons field ($cb$).  Reducing the degeneracy between $M_{\nu}$ and $\sigma_8$ is achieved via Persistent Homology for the $m$ field but not for the $cb$ field.  The uncertainty of $M_{\nu}$ at $1\sigma$ confidence interval from the joint analysis of Persistent Homology vectorization for the $m$ and $cb$ fields smoothed by $R=5$ Mpc $h^{-1}$ at $z=0$ reaches $0.0152$ eV and $0.1242$ eV, respectively. Noticing the use of the 3-dimensional underlying density field at $z=0$, the mentioned uncertainties can be treated as the theoretical lower limits.

    \end{abstract}




\section{Introduction}
Despite the success of the standard cosmological model, vanilla-$\Lambda$CDM, (among many for example see \cite{aghanim2020planckvi,eBOSS:2020yzd,Brout:2022vxf,46ddc1eeccc74c52ba263dcebe509c3a}), it is necessary to well constrain the associated extensions in the high-precision experiments \citep{EUCLID:2011zbd,LSSTScience:2009jmu,SimonsObservatory:2018koc,CMB-S4:2016ple}. From an observational perspective, we anticipate an outstanding opportunity for a comprehensive exploration of galaxy clustering through current and upcoming surveys. These surveys include DESI\footnote{\texttt{http://www.desi.lbl.gov}}, PFS\footnote{\texttt{http://pfs.ipmu.jp}}, the Roman Space Telescope\footnote{\texttt{http://wfirst.gsfc.nasa.gov}}, Euclid\footnote{\texttt{http://sci.esa.int/euclid}}, and CSST\footnote{\texttt{http://nao.cas.cn/csst}} \citep{2011SSPMA,2019ApJ883}. In addition, advancements in CMB experiments, particularly the next generation of observations such as CMB Stage IV \citep{2019arXiv190704473A,2017arXiv170602464A}, are paving the way for a more precise evaluation of extensions to the concordance cosmological model and their associated exotic features. Furthermore, these advancements present possibilities for future enhancements. Simulations, a critical component of scientific methodology, also provide high-precision evaluations and deeper insights into more complex N-body systems and their underlying motivations. One significant extension to the standard cosmological model is the $\nu \rm \Lambda CDM$ cosmology, which considers the sum of neutrino masses as a free parameter beyond the minimal value \citep{Dodelson:1995es,2009PhRvL.103q1301D,lesgourgues2013neutrino,Euclid:2024imf}. From the particle physics point of view, determining the mass of neutrinos is highly motivated \citep{Abazajian:2017tcc,Capozzi:2018ubv}. The mass-squared differences are well determined in three-flavor oscillation analyses of experiments such as reactor, atmospheric, and solar neutrino observations \citep{Esteban:2020cvm}\footnote{\texttt{http://www.nu-fit.org/}}. However, the absolute mass scale of neutrinos remains unknown. The lower bound on the sum of neutrino mass from oscillation experiments is $\sum_\nu m_\nu \ge 0.06 \ \rm eV$ for normal and $\sum_\nu m_\nu \ge 0.1 \ \rm eV$ for inverted hierarchy \citep{Capozzi:2017ipn,Kelly:2020fkv}. The upper limits on the effective electron anti-neutrino mass has been narrowed down to $ m_{\beta}<0.8 \ \rm eV$ at $90\% \ {\rm confidence\ level\ (CL)}$ from Karlsruhe Tritium Neutrino experiment, which is equivalent to $\sum m_{\nu}\lesssim 2.4$ eV at $90\%$ CL.

In cosmology, cosmic neutrinos with non-zero mass not only change the expansion rate of the Universe but also impact the evolution of perturbations. \citep{Lesgourgues:2006nd,2011ARNPS..61...69W}.
 When massive neutrinos transition to a non-relativistic state, they exhibit behavior similar to cold dark matter. Consequently, they contribute to the clustering of matter on scales larger than their free-streaming scale ($\lambda_{fs}$). The $\lambda_{fs}$ represents the characteristic length scale over which perturbations in massive neutrinos propagate. Additionally, massive neutrinos suppress the growth of structures on scales smaller than $\lambda_{fs}$ \citep{Lesgourgues:2006nd,lesgourgues2013neutrino}. Therefore, the absolute mass of neutrinos significantly influences the distribution of matter density, the growth rate of structures, and the overall background cosmological evolution. Massive neutrinos also modify the shape of the matter power spectrum, the clustering of halos in both real and redshift space, and the number counts of critical structures such as voids. Additionally, they affect the dependence of the bias factor, as discussed in sources such as \cite[and references therein]{Contarini2020CosmicVI,Moon2023ApJS}.

 Assuming a $\Lambda$CDM model, data from the Cosmic Microwave Background (CMB), including temperature and polarization measurements, places an upper limit on the mass of massive neutrinos at approximately $0.26 \ \rm eV$ with $95\%$ confidence level (CL) \citep{aghanim2020planckvi}. Numerous cosmological observations have attempted to constrain neutrino mass by analyzing the shape and amplitude of the matter power spectrum. However, due to the degeneracy between neutrino mass and other cosmological parameters, the results have not been distinctive or highly promising so far. A more stringent upper bound on the mass of massive neutrinos can be obtained by combining the CMB primary power spectrum with Baryon Acoustic Oscillations (BAO) and CMB lensing data, yielding $\sum_\nu m_\nu \le 0.12 \ \rm eV$ at $95\%$ CL \citep{2017PhRvD..96l3503V,2018PhRvD..98h3501V,aghanim2020planckvi,2022JHEAp..36....1T}. When SPT-3G CMB anisotropy, SPT-3G lensing data, and BAO data are combined, the upper limit is $M_{\nu} < 0.30$ eV at the 95\% confidence level \citep{SPT:2023jql}. Similarly, combining observations from the Atacama Cosmology Telescope (ACT) from 2017 to 2021 with Planck lensing data establishes a limit of $\sum m_{\nu}<0.13$ eV at the 95\% CL \citep{ACT:2023kun}. Additionally, a new constraint on the total neutrino mass is provided by combining DESI and CMB data, resulting in $\sum m_{\nu}<0.072$ eV at the 95\% CL for the $\Lambda$CDM model \citep{2024arXiv240403002D}. The Kilo-Degree Survey (KiDS-1000) reports a marginal constraint of $\sum m_{\nu}<1.76$ eV at 95\% CL \citep{KiDS:2020ghu}.
 
Numerous efforts have been made to clarify various aspects of cosmological implications in neutrino physics beyond the reported constraints on massive neutrinos. These efforts include investigating the influence of cosmological neutrinos on the dark matter halo mass function, clustering properties, and redshift space distortions \citep{2011MNRAS.418..346M}. Researchers have also examined the effect of massive neutrinos on the clustering of dark matter halos within the context of scale-dependent bias \citep{2014JCAP...03..011V,2014JCAP...02..049C,Chiang2017ScaledependentBA,Chiang2018FirstDO} and halo assembly bias \citep{2021JCAP...03..022L}. Additionally, alternative statistical methods, such as Minkowski Functionals \citep{liu2020neutrino,Liu2022ProbingMN,Liu2023ProbingMN}, marked power spectrum \citep{2021PhRvL.126a1301M}, bispectrum \citep{2020JCAP...03..040H}, and counts-in-cells statistics \citep{2020MNRAS.495.4006U}, have been employed to probe massive neutrinos. The use of geometric features to analyze massive neutrinos has also been explored \citep{2019JCAP...06..019M,2021JCAP...01..009P,Moon2023ApJS}. Further, the impact of massive neutrinos on cosmic voids has been investigated \citep{2015JCAP...11..018M,2019MNRAS.488.4413K,Schuster2019TheBO,Vielzeuf2023DEMNUniTI}, and modified gravity effects with massive neutrinos on cosmic voids have been assessed \citep{Contarini2020CosmicVI}. Efforts have also been made to reduce the degeneracy among cosmological parameters, particularly those involving massive neutrinos \citep[and references therein]{2020JCAP...03..040H,2021ApJ...919...24B}. The significance of massive neutrinos in non-linear clustering of matter fields has been extensively studied in both real-space \citep{2011MNRAS.410.1647A,2016JCAP...11..015B} and redshift-space \citep{2016PhRvD..93f3515U}, with further details available in \citep[and references therein]{2020JCAP...03..040H}.

The interplay between the parameters of the matter power spectrum and the inclusion of neutrino mass as an additional free parameter can weaken the analysis from the perspective of massive neutrino cosmology. The two-point statistics of LSS on small scales do not fully capture the imprint of massive neutrinos. However, it has been demonstrated that for an almost 2-dimensional density field in real space and up to $k\lesssim‌1$ Mpc$^{-1}h$, the power spectrum can almost figure out the influence of massive neutrinos, while for a 3-dimensional field and in redshift space, the non-linear scales contain much information from massive neutrinos \citep{2022PhRvD.105l3510B}. Utilizing higher-order spectra such as the bispectrum is capable of placing more stringent constraints on the total mass of massive neutrinos \citep{2019JCAP...11..034C,2020JCAP...03..040H}.

 In this context, developing new estimators that can extract more detailed information about the neutrino mass $M_{\nu}$ from observations of matter density is both crucial and promising. Additionally, introducing new statistical measures that probe various scales could yield more robust constraints on $M_{\nu}$, given the scale-dependence of matter density when massive neutrinos are present.

The emergence of tensions and degeneracies among different cosmological parameters has driven researchers to employ more complex algorithms for the systematic evaluation of high-dimensional data. The intersection of algebraic topology and computational geometry offers an opportunity to enhance our understanding and provide deep insights into the quantitative assessment of various cosmological fields \citep{2024arXiv240313985Y}. Topological characteristics of a manifold describe properties that are preserved under continuous deformations. Additionally, topology encompasses aspects such as size, shape, connectedness, and boundaries. The cosmological mass distribution field has been analyzed by focusing on topological invariants such as the genus and the Euler characteristic \citep{1986ApJ...306..341G,1986ApJ...309....1H,1989ApJ...340..625G,pranav2019topology,2021MNRAS.507.2968W}. The application of topological invariants, including Euler characteristics and Minkowski Functionals (MFs), as well as their extensions known as Minkowski Valuations, has provided deeper insights into cosmological implications \citep{McMullen1997,Alesker1999,Beisbart2002,matsubara2003statistics,Hug2007TheSO,2019A&A...627A.163P,matsubara2022minkowski,2024ApJ...963...31K}. Recent interest has focused on using simplicial complexes to assign shapes to discrete and continuous high-dimensional data and computing various dimensional holes for given proximity parameters (thresholds). This approach has garnered attention due to its ability to capture more detailed information \citep[and references therein]{2018JCAP...03..025C,2021MNRAS.507.2968W,2021A&A...648A..74H,2022A&A...667A.125H,2023MNRAS.522.2697T,2022JCAP...10..002B}. Persistent Homology (PH), a method within Topological Data Analysis (TDA) \citep{dey2022computational,edelsbrunner2022computational,zomorodian2005topology11,wasserman2018topological}, offers a novel approach by utilizing topological invariants to extract information from complex data sets, revealing distinct aspects of the underlying structures \citep{otter2017roadmap,pereira2015persistent,masoomy2021persistent,2022PhRvE.106f4115M}. PH is a mathematical tool used to analyze and understand large-scale structures LSS by detecting and tracking the persistence of topological features such as clusters, filament loops, and voids across different scales \citep{van2011,Xu2019,2021MNRAS.507.2968W,biagetti2021persistence,2023MNRAS.522.2697T,2023MNRAS.520.2709E}. The use of PH to differentiate between hot and cold dark matter models has been explored by \cite{Cisewski-Kehe2022PhRvD}. Additionally, PH has been employed to investigate non-Gaussianity from a perspective different from conventional statistical methods \citep{feldbrugge2019stochastic,biagetti2021persistence,2022JCAP...10..002B}. Besides topological measures, interesting geometric criteria have also been applied to cosmological analysis, particularly to place precise constraints on cosmological parameters, including the total mass of neutrinos \citep{2023PhRvX..13a1038P}. Recently, multiscale topological characteristics of LSS, along with joint analyses of power spectrum and bispectrum statistics to constrain cosmological parameters, have been reported in \cite{2024arXiv240313985Y}.

Pioneering research by \cite{2022AAS...24031208R} has highlighted the influence of topological properties of the cosmic web, particularly its filamentary structure, on the presence of massive neutrinos. This study specifically investigated filament-based statistics and persistence diagrams within the context of cosmologies involving massive neutrinos. The results indicated that the cosmic web at high redshifts is significantly impacted by the presence of massive neutrinos.

The primary impact of massive neutrinos on the  LSS is related to the sum of the masses of the three neutrino species ($M_{\nu}$). Therefore, in this paper, we focus on $M_{\nu}$ as our case study. We examine the almost nonlinear regime at the present cosmic time, $z=0$, where massive neutrinos have a significant influence on the LSS \citep{Vielzeuf2023DEMNUniTI}. At higher redshifts and in the linear regime, massive neutrinos, particularly at large scales when they become non-relativistic, behave similarly to CDM. In these cases, both weighted and unweighted Two-Point Correlation Functions (TPCF) and measures based on the power spectrum can serve as appropriate criteria for evaluation \citep{2021JCAP...01..009P,2021MNRAS.503..815V}.
To reduce degeneracies and address discrepancies in parameter constraints, higher-order spectrum analysis should be considered for low redshifts \citep{2020JCAP...03..040H}. In this paper, we investigate the subtle effects of massive neutrinos on the LSS using components of PH within an algebraic-topological framework \citep{nakahara2003geometry,munkres2018elements}. The novelties and advantages of our current research are outlined below:\\
(1) We will provide a comprehensive constraint on cosmological parameters, including $M_{\nu}$, using PH vectorization for the total matter density field ($m$) and for cold dark matter plus baryons ($cb$) in simulations of massive neutrinos from the Quijote suite.\\
(2) We evaluate and highlight the potential of PH for probing the presence of massive neutrinos and reducing degeneracies between various cosmological parameters. Specifically, we focus on the degeneracy in the $(M_{\nu}, \sigma_8)$ plane using fully nonlinear measures and the cosmic web at $z=0$. This proof of concept demonstrates that PH vectorization can be a valuable tool for cosmological inferences.\\
(3) We introduce an efficient estimator inspired by the moments of a typical field and quantitatively assess the sensitivity of different topological invariants to neutrino mass, along with their cosmological interpretations. Additionally, we examine the impact of the smoothing scale on our results.\\

The rest of paper is organized as follows. Section \ref{sec:ph} reviews the concept of homology in the context of the excursion set theory and its application to discrete cosmological datasets.
In section \ref{sec:data}, we present the description of the simulations.
Section \ref{sec:imp} is devoted to probing the signature of massive neutrinos on the simulated LSS by Quijote suite.
In section \ref{sec:fish}, we quantify the information content of PH to constrain the $\nu\Lambda$CDM parameters and total neutrino mass using Fisher analysis formalism.  We will give our summary and conclusion in section \ref{sec:con}.
\begin{figure*}
	\includegraphics[width=2\columnwidth]{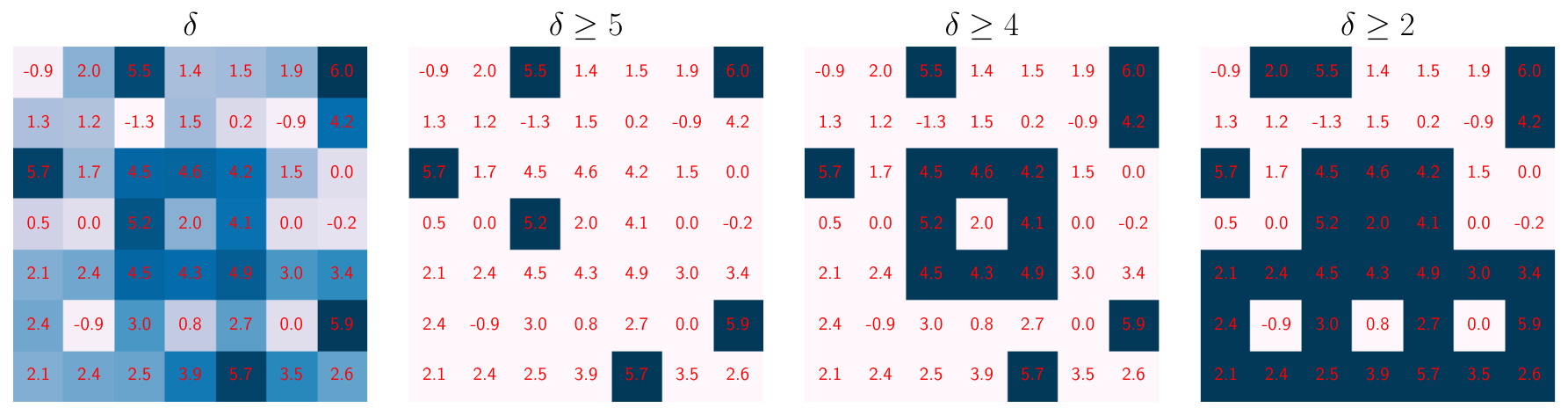}
	\caption{Evolution of topological features by changing the filtration parameter (known as proximity parameter) in the context of super-level filtration for a 2-dimensional mock density field. The left panel shows the density field and the number (color) assigned to each cell indicates the value of density contrast ($\delta$).  The sub-complexes corresponding to the proximity level $\vartheta= \{5,4,2\}$ in such that $\delta (\boldsymbol{r})\ge \vartheta$ are indicated by assigning the dark color for pixels in the second, third, and fourth panels from the left to right, respectively.}\label{fig:fig1}
\end{figure*}
\section{The Foundation tenets  of PERSISTENT HOMOLOGY}\label{sec:ph}
In this section, we outline pertinent features of PH for quantifying the morphology of discrete data sets.
Consider a cosmological dataset sampled on a cubical lattice, $\mathcal{M}$. This dataset may consist of the locations of dark matter particles (or dark matter halos) obtained from N-body simulations, or the positions of galaxies from a cosmic survey. For this work, we focus on the positions of dark matter particles. Assume that the lattice $\mathcal{M}$ is divided into $N^3_{grid}$ regular cubic cells. Given the positions of the dark matter particles at a specific time, we assign a number density contrast to each cell of the lattice as follows:
\begin{eqnarray}
	\delta(\boldsymbol{r},t) \equiv \frac{n (\boldsymbol{r},t)-\bar{n}(t)}{\bar{n}(t)}
	\label{eq:density_field}
\end{eqnarray}
where $n (\boldsymbol{r},t)$ represents the number of particles placed in the cell associated with the location $\boldsymbol{r}$ at cosmic epoch, $t$, and $\bar{n}(t)$ shows the mean number of particles over all cells at $t$. According to the definition given by Eq. (\ref{eq:density_field}), we have $\delta\in[-1,+\infty)$. Throughout this paper, we fix ``$t$" to the present epoch, taking a snapshot of the simulation at the final evolution epoch, $z=0$. Consequently, we omit the time-dependency of our introduced parameters. Our current research does not address the redshift dependency, which we leave for future work. 

Mathematically, the lattice  $\mathcal{M}$ can be considered as a topological space or more precisely a cubical complex, and  $\delta$ can be interpreted as a mapping from the topological space to the real space, indicated by  $\delta: \mathcal{M} \rightarrow \mathbb{R}$, which $\mathcal{M} \subset \mathbb{R}^3$. Applying the definition of the excursion sets on $\mathcal{M}$ via the super-level filtration  leads to the subset  $\mathcal{M}_{\vartheta} = \left\{ \boldsymbol{r} \in \mathcal{M}\; |\; \delta(\boldsymbol{r}) \ge \vartheta\right\}$ associated with threshold, $\vartheta$. While this work utilizes super-level filtration, it is important to note that in general, the choice of filtration should be adapted to the specific type of dataset (e.g. see \citep{2024arXiv240313985Y}). By varying the threshold from a maximum value to a minimum value, $\vartheta_{max} \geq \dots \geq \vartheta_{i+1} \geq \vartheta_{i} \geq \dots \geq \vartheta_{min}$, one can obtain a sequence of subsets of $\mathcal{M}$ as:
\begin{eqnarray}
	\mathcal{M}_{\vartheta_{max}} \subseteq \dots \mathcal{M}_{\vartheta_{i+1}} \subseteq 
	\mathcal{M}_{\vartheta_{i}} \subseteq \dots \mathcal{M}_{\vartheta_{min}} 
\end{eqnarray}
If $\mathbb{M} = \left\{\mathcal{M}_{\vartheta_{i}}\right\}_{i}$ implies the set of subspaces, the main purpose of Topological Data Analysis, or  more specifically Persistent Homology, can be expressed as extracting the topological information (topological invariants) from $\mathbb{M}$.

In algebraic topology, a topological invariant refers to a property of a topological space that remains unchanged under continuous deformations such as stretching, shrinking, reflecting, or rotating. Practically, for our typical topological space $(\mathcal{M}_{\vartheta_i})$ within the context of classical topology, the homology of $\mathcal{M}_{\vartheta_i}$ characterizes its topological properties. For example, if $\mathcal{M}_{\vartheta_i}$ is a 3-dimensional embedded cubical complex, it has three homology groups:
$H_{k}(\mathcal{M}_{\vartheta_i})$ with $k = 0,1,2$. The Betti numbers, $\tilde{\beta}_k$, which are topological invariants, can be determined by the ranks of homology groups: $\tilde{\beta}_k (\mathcal{M}_{\vartheta_i}) = \big|H_k(\mathcal{M}_{\vartheta_i})\big |$. The elements of the $k$th homology group correspond to the $k$-dimensional holes ($k$-holes) of the cubical complex. Therefore, the Betti numbers provide information about the number of connected components, independent loops, and independent closed surfaces in a topological space. 

The filtration procedure constructs a set of (sub-)cubical complexes ($\mathbb{M}$) from a discrete dataset. One straightforward approach to studying the set $\mathbb{M}$ is to examine the topological properties of its sub-complexes through their homology groups $H_k(\mathcal{M}_{\vartheta_i})$ individually. PH goes further by quantifying the evolution of topological features as the filtration parameter changes, and by examining the inclusion maps $\mathcal{M}_{\vartheta_{i}} \rightarrow \mathcal{M}_{\vartheta_{i+1}}$. This evolution is characterized by the appearance ($birth$) and disappearance ($death$) of topological features. The PH of $\mathbb{M}$ is denoted by $H_k(\mathbb{M})$, which provides information in the form of persistent pairs $\vartheta^{(k)} = \left( \vartheta^{(k)}_{birth}, \vartheta^{(k)}_{death}\right)$. A persistent pair indicates that a $k$-hole appears at level $\vartheta^{(k)}_{birth}$ and disappears at level $\vartheta^{(k)}_{death}$. The information from PH can be summarized in a multiset $\mathcal{D}_k = \left\{\vartheta^{(k)}_i \right\}_i$,  which is a collection of all persistence pairs and is known as the persistence diagram. In Fig. \ref{fig:fig1}, we illustrate the density contrast field of a 2-dimensional synthetic dataset. The field is sampled within the range $-1 \le \delta \le 6$, and some of its sub-complexes obtained through filtration are visualized. Cells that satisfy the super-level condition are shown in dark color.  As the proximity parameter decreases, $k$-holes appear, and at certain thresholds, they merge and subsequently disappear. We can also define:
\begin{eqnarray}
	\vartheta^{(k)}_{(i),pers} \equiv \left |\vartheta^{(k)}_{(i),birth} - \vartheta^{(k)}_{(i),death}\right|
\end{eqnarray}
to measure the persistency of a pair. Therefore one can construct the persistence diagram in terms of $\vartheta^{(k)}_{(i),birth}$, $\vartheta^{(k)}_{(i),death}$ and, $\vartheta^{(k)}_{(i),pers}$. Although the multiset $\mathcal{D}_k$  contains the PH information and can be used for demonstration purposes, but to achieve quantitative results for further analysis, we use statistical criteria defined over the persistence pairs \citep{biagetti2021persistence,masoomy2021persistent,2022PhRvE.106f4115M}. The persistence diagram is the raw output of PH. To construct tractable and interpretable topological features that are sensitive to cosmological parameters and suitable for cosmological inferences (e.g., Fisher forecast analysis), various methods have been introduced to map the persistence diagram to summary statistics \citep{adams2017persistence}. Additionally, computing the associated covariance matrix from a persistence diagram can be time-consuming due to challenges related to high dimensionality. Therefore, in the remainder of this section, we will provide a brief review of some one-dimensional summary statistics.

We begin with the well-known Betti curves, which are collections of all Betti numbers. These curves are extracted from sequences of sub-complexes and sorted by the filtration parameter. Betti curves visualize the evolution of $k$-hole populations as a function of $\vartheta$. For a fixed $\vartheta$, the Betti number $\tilde{\beta}_k(\vartheta)$ represents the number of pairs that satisfy the conditions $\vartheta^{(k)}_{(i),birth}\geq \vartheta$ and $\vartheta^{(k)}_{(i),death}\le \vartheta$. Mathematically, this can be written as: 
\begin{eqnarray}
\tilde{\beta}_k(\vartheta) = \sum_{i = 1}^{n_k} \Theta \left(\vartheta^{(k)}_{(i),birth} - \vartheta\right) \Theta \left(\vartheta - \vartheta^{(k)}_{(i),death}\right)
\end{eqnarray}
where $\Theta(:)$ indicates Heaviside function, and $n_k$  represents the total number of persistence pairs related to the $k$th persistence diagram. Throughout this paper, we normalize the Betti numbers $\tilde{\beta}_k$ by the simulation volume, and refer to them as the normalized Betti numbers ($\beta_k$). It is important to note that Betti curves do not capture all the information encoded in persistence diagrams. To provide complementary assessments, we will use two additional types of visualizations proposed by \cite{biagetti2021persistence}.
These visualizations represent the number of persistence pairs satisfying specific conditions: $\vartheta^{(k)}_{(i),birth}\geq \vartheta$ and $\vartheta^{(k)}_{(i),pers}\geq \vartheta$ for $\tilde{B}_k(\vartheta)$ and $\tilde{P}_k(\vartheta)$, respectively. The mathematical form of $\tilde{B}_k(\vartheta)$ and $\tilde{P}_k(\vartheta)$ become:
\begin{eqnarray}
	\tilde{B}_k(\vartheta) \equiv  \sum_{i = 1}^{n_k} \Theta \left(\vartheta^{(k)}_{(i),birth} - \vartheta \right)
\end{eqnarray}
and 
\begin{eqnarray}
	\tilde{P}_k(\vartheta) \equiv \sum_{i = 1}^{n_k} \Theta  \left(\vartheta^{(k)}_{(i),pers} - \vartheta\right)
\end{eqnarray}
To report the $\tilde{B}_k$ and $\tilde{P}_k$ for underlying data sets, we use the empirical distribution function as the numerical estimator for the mentioned topological measures denoted by $B_k(\vartheta)$ and $P_k(\vartheta)$. In the next section, we will focus on the quantitative measures of topological properties. Based on our pipeline, we will implement the PH vectorization on mock data for cosmological inferences. 
\begin{figure}
	\centering
	\includegraphics[width=1\columnwidth]{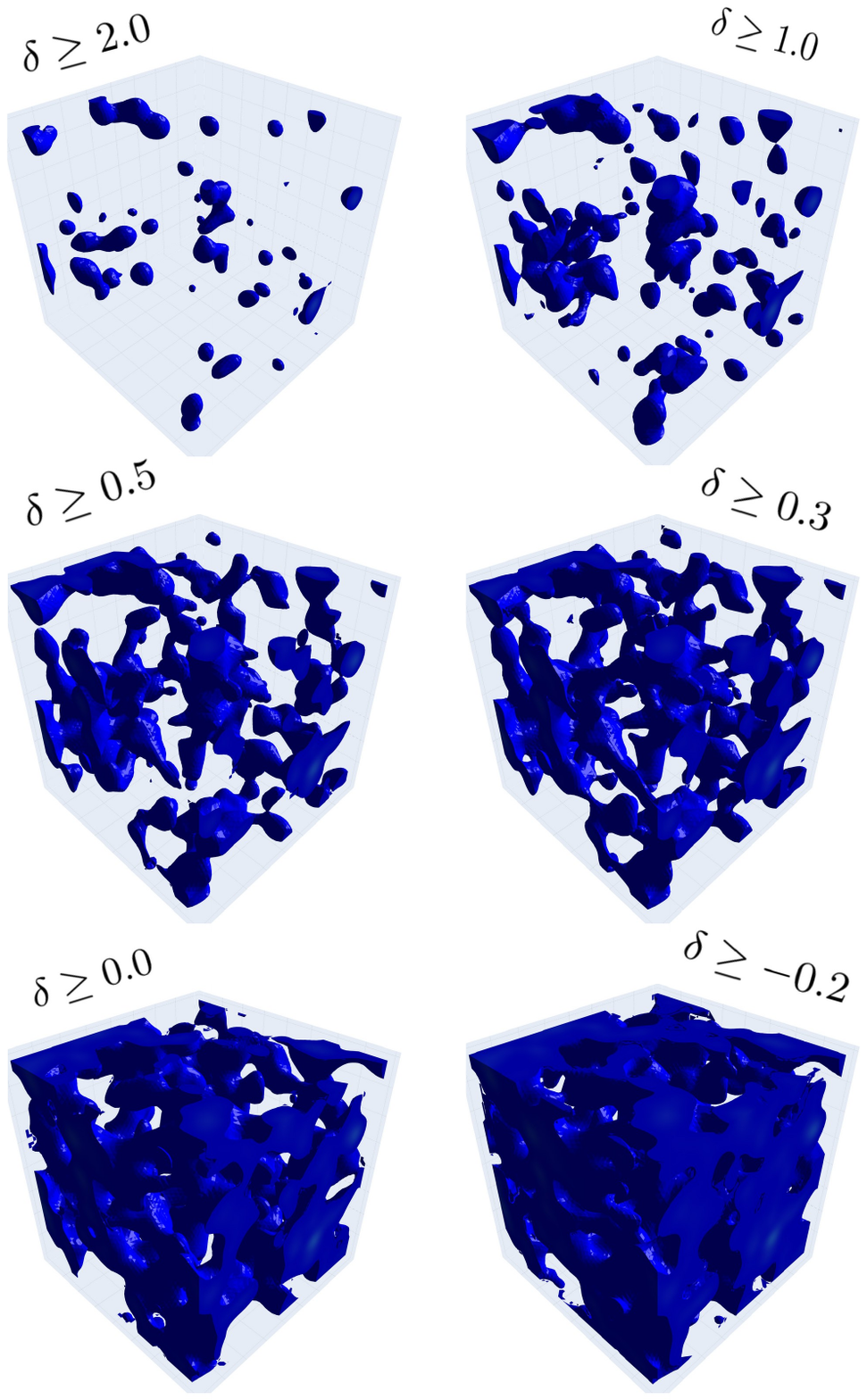}
	\caption{Excursion sets were made with super-level filtration on the density field for a fiducial realization from Quijote simulations. We cropped box of $156$ Mpc $h^{-1}$ size from the original volume. The density field is constructed based on the particle position at $z=0$ using the cloud-in-cell scheme performed by   \texttt{Pylians}. Here we consider six threshold levels,  $\vartheta =\{2.0, 1.0, 0.5, 0.3, 0.0, -0.2\}$ in such that $\delta(\boldsymbol{r},z=0)\ge \vartheta$. To smooth the constructed density field, we use the Gaussian window function with smoothing scale $R=5$ Mpc $h^{-1}$.  }
	\label{fig:F22}
\end{figure}

\begin{figure*}
	\centering
	\includegraphics[width=1.5\columnwidth]{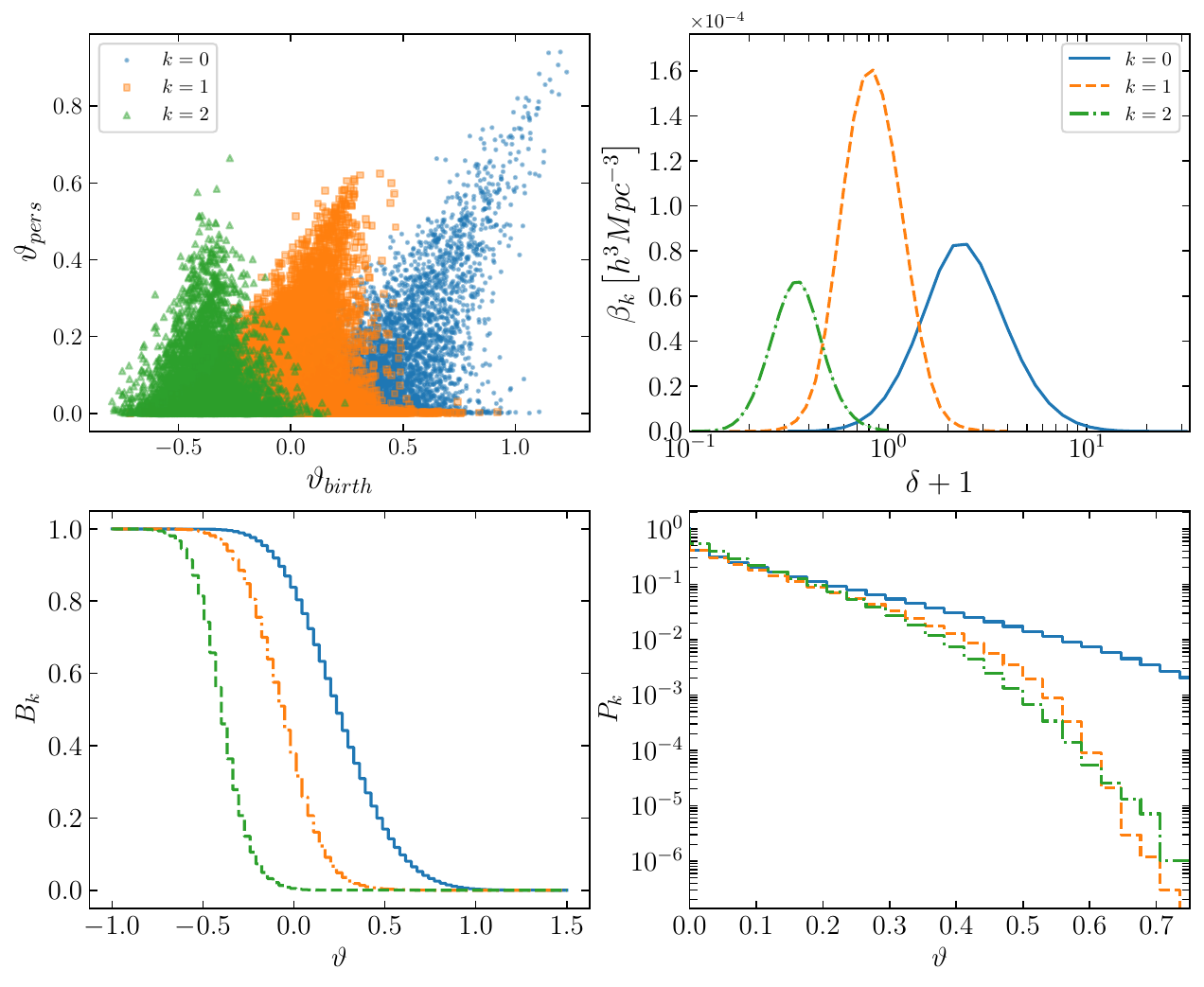}
	\caption{The extracted persistence diagram of a fiducial realization from Quijote simulations. The upper left panel reveals the persistence diagram in the scatter plot for $0$-, $1$- and $2$-holes. The $\beta_k$ as a function of density threshold ($1+\delta$) is represented in the upper right panel. The complementary representations of the persistence diagram, namely $B_k$ and $P_k$ are indicated in the lower left and lower right panels, respectively. It is worth mentioning that the $\vartheta$ for the $B_k$ and $P_k$ quantifies respectively the birth and persistency thresholds.}
	\label{fig:F2}
\end{figure*}

\section{Data description}\label{sec:data}
In this study, we investigate the imprints of massive neutrinos on the large-scale structure (LSS) using Persistent Homology (PH). We utilize a subset of simulations from the Quijote suite \citep{Quijote_sims}, which comprises 44,100 N-body realizations. The TreePM code GADGET-III is used to generate all realizations within the Quijote simulations. Each simulation in this work consists of $512^3$ gravitationally interacting dark matter particles, and for simulations involving massive neutrinos, there are also $512^3$ neutrino particles. These particles are sampled within a cubic volume of $1\ \text{Gpc}^3\ h^{-3}$ with periodic boundary conditions. The redshift interval for the simulation is $z\in[0,127]$. To generate the initial conditions for both massive neutrino simulations and the fiducial massless counterparts, we apply the rescaling method  \citep{2017MNRAS.466.3244Z} and Zel'dovich approximation are applied. For all other simulations, second-order perturbation theory is used.
The cosmological parameters related to the fiducial simulations are based on the flat $\Lambda$CDM model including $\Omega_m=0.3175$, $\Omega_b=0.049$, $h=0.6711$, $n_s=0.9624$ and $\sigma_8=0.834$ which are in a good agreement with {\it Planck} observations \citep{aghanim2020planckvi}.

To examine the various aspects of massive neutrinos, it is essential to consider the total matter field, which includes CDM, baryons, and massive neutrinos. Additionally, because massive neutrinos influence the evolution of the CDM and baryon fields, we can focus on the combined CDM and baryon field in simulations to study the effects. Henceforth, we refer to the total matter field as $``m"$ and the combined CDM and baryon field as $``cb"$. It is important to note that the bias factor plays a crucial role in translating theoretical predictions based on the total mass into observational quantities measured by various surveys. It has been demonstrated that when examining the impact of massive neutrinos on the clustering of dark matter halos, without considering the clustering of the massive neutrinos themselves, the combined CDM and baryon field (denoted as ``$cb$") is the relevant quantity \citep{2012PhRvD..85f3521I,2014JCAP...02..049C}. To assess the impact of massive neutrinos on the LSS, the combined CDM and baryon field is the most relevant tracer we observe \citep{2001MNRAS.321..372J,2003MNRAS.346..565R}. Subsequently, we will evaluate the various components of the total matter field in relation to the massive neutrinos. The density contrast for the total matter field, based on the classifications mentioned, is defined as follows::
\begin{eqnarray}
	\delta_{m}\equiv\frac{\Omega_{CDM}}{\Omega_{m}}\delta_{CDM}+\frac{\Omega_{b}}{\Omega_{m}}\delta_{b}+\frac{\Omega_{\nu}}{\Omega_{m}}\delta_{\nu}
\end{eqnarray}
where $\Omega_{m}=\Omega_{CDM}+\Omega_{b}+\Omega_{\nu}$. 
Also, for the Quijote simulations, one can assume $\delta_{CDM} = \delta_{b} \equiv \delta_{cb} $. The $\delta_{cb}$ and $\delta_{\nu}$ correspond to the density contrast of CDM+baryons and neutrinos, respectively. Therefore, we have $\delta_m = \frac{\Omega_{cb}}{\Omega_{m}}\delta_{cb} + \frac{\Omega_{\nu}}{\Omega_{m}}\delta_{\nu}$. The motivation for selecting the mentioned fields for cosmological applications will be explained in the next section. The total mass of neutrino particles in the Quijote simulations for massive neutrinos is categorized into $M_{\nu}^{+} = 0.1$ eV, $M_{\nu}^{++} = 0.2$ eV, and $M_{\nu}^{+++} = 0.3$ eV.  

We also utilize the \texttt{Pylians} \citep{villaescusa2018pylians} routines to construct density contrast fields from Quijote snapshots for both $m$ and $cb$ fields. We adopt a cloud-in-cell (CIC) mass-assignment scheme to construct fields on regular grid spacing. The presence of threshold-based analysis in our pipeline essentially leads to considering the smoothed field. The smoothed field is produced by the convolution of the underlying field with a smoothing window function, which is intrinsically based on the distance from the center of each cell \citep{1994A&A...291..697B,2007PASJ...59...73M}. We apply a Gaussian kernel with two characteristic smoothing scales $R = 5$  Mpc\; $h^{-1}$ and  $R=10$ Mpc $h^{-1}$. It is important to note that the massive neutrino simulations used in this paper contain $512^3$ neutrino particles. As a result, shot noise is expected due to the discrete nature of these simulations. We have verified that the adopted smoothing scales, which are larger than the voxel size associated with the $N_{grid}=512^3$, effectively mitigate the impact of shot noise on our results. Taking a larger smoothing scale reduces the sensitivity of our analysis to higher thresholds of the density field as well as to non-linearities. The primary source of these non-linearities is the gravitational growth of structures, which is influenced by massive neutrinos. In other words, massive neutrinos inhibit the growth of structures on scales smaller than their free-streaming length scale ($\lambda_{fs}$). On scales larger than $\lambda_{fs}$, massive neutrinos behave similarly to CDM particles. The $\lambda_{fs}$ dependency of massive neutrinos introduces scale-dependent effects on the clustering of structures in both linear and non-linear regimes, which can be probed by varying the smoothing scale. The $\lambda_{fs}$ itself depends on the neutrino mass $M_{\nu}$ and redshift \citep{lesgourgues2013neutrino,2012arXiv1212.6154L}. At redshift $z = z_{{\rm nr}}$, when massive neutrinos become non-relativistic, $\lambda_{fs}$ reaches its minimum value. At all redshifts smaller than $z_{{\rm nr}}$, the matter field is influenced by massive neutrinos in both linear and non-linear regimes \citep{2012arXiv1212.6154L}. Although it would be interesting to investigate how the effect of neutrino mass on the LSS changes with redshift, we focus on a snapshot of the field evolution at its final epoch ($z = 0$), when non-linear effects are most pronounced.

As an illustration, Fig. \ref{fig:F22} displays the excursion sets using the super-level filtration for the constructed matter density field of a fiducial realization from the Quijote simulations at various threshold levels. The matter density field has been smoothed using a Gaussian window function with a smoothing scale $R=5$  Mpc $h^{-1}$.

\begin{figure*}
	\centering
	\includegraphics[width=2.1\columnwidth]{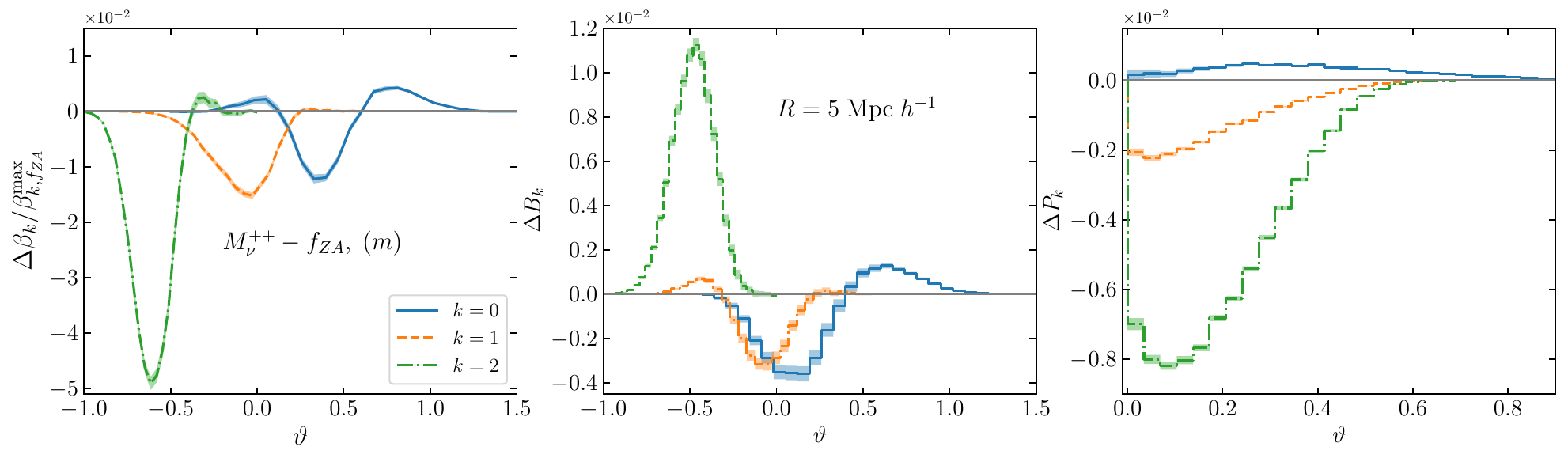}
	\includegraphics[width=2.1\columnwidth]{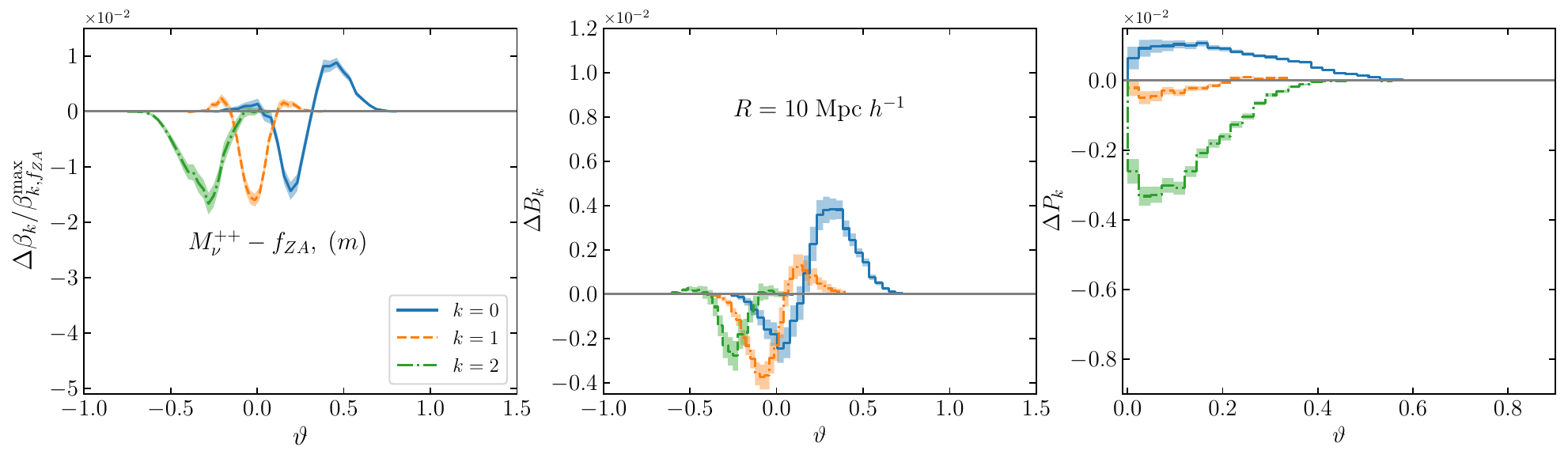}
	\caption{The PH vectorization for the $m$ field when the massive neutrinos particles with the total mass $M^{++}_{\nu} = 0.2 \ \rm eV$ are added compared to the fiducial cosmology. The left panels represents the Betti curves divided by the corresponding maximum value in the fiducial case. The middle and right panels are devoted to the differences in the $(B_k,P_k)$ with respect to fiducial cosmology, respectively. The blue solid, orange dashed and green dashed-dotted lines represent the 0-hole, 1-hole, and 2-hole homology groups, respectively. The shaded areas are associated with the $2\sigma$ confidence interval errors estimated over $200$ realizations. The upper and lower rows are devoted to smoothing scales  $R=5$ Mpc $h^{-1}$ and $R=10$ Mpc $h^{-1}$, respectively.  }
	\label{fig:F3_new}
\end{figure*}

\section{Implementation of PH on the Quijote N-body simulations}\label{sec:imp}
In this section, we present the numerical derivation of the topology of a subset of Quijote simulations across different cosmological models using our PH vectorization, $(\beta_k,B_k,P_k)$. Of particular interest in topology terminology is the quantification of topological features, such as islands, loops, shells, and higher-dimensional shapes, based on the dimension of the underlying field.
Subsequently, for cosmic structures in the 3-dimensions, we compute $\beta_0$ (the number of components), $\beta_1$ (the number of independent loops), and $\beta_2$ (the number of independent closed surfaces) (for more details see \citep{2017MNRAS.465.4281P,pranav2019topology,2019A&A...627A.163P,edelsbrunner2022computational,2021MNRAS.507.2968W}). 	    
To make a correspondence between Betti numbers and cosmological inference, recently, the so-called Significant Cosmic Holes in Universe method has been introduced for identifying different components of cosmic web in  \citep{Xu2019}.  Recognizing voids, filaments, and clusters depends not only on how the underlying matter fields are tessellated but also on the procedure used to evaluate robustness against the filtration process and quantify the reliability of persistent pairs in the persistence diagram. This procedure has a significant impact. In summary, one should exercise caution when comparing the effects of total neutrino mass on derived components using a topological-based approach with common methods used for identifying voids \citep{2015JCAP...11..018M,2017MNRAS.465.4281P,2019MNRAS.488.4413K,pranav2019topology,Xu2019,Contarini2020CosmicVI}. In practice, when observing 2-dimensional and pseudo-2-dimensional fields, such as lensing maps and projected maps of tracers, the uncertainty level of PH vectorization extracted from the 3-dimensional underlying density field can be regarded as the theoretical lower (upper) bound on the uncertainty (accuracy) level.

\subsection{PH of fiducial vanilla-$\Lambda$CDM  cosmology }

For clarity, and before discussing how variations in cosmological parameters affect the topological criteria, we present the persistence diagram and associated vectorization for a realization of the fiducial simulations in Fig.~\ref{fig:F2}. The upper left panel illustrates the persistence diagram in the scatter plot  in the birth-persist coordinate. Each of the blue circles, orange squares and green triangles represents
a persistence pair $(\vartheta^{(k)}_{(i),birth},\vartheta^{(k)}_{(i),pers})$ corresponding to the 0-, 1- and 2-holes, respectively. 

As illustrated by Fig.~\ref{fig:F2},  the 0-, 1- and 2-holes mainly appear at high, middle, and low thresholds, respectively. This behavior can be explained by the nature of connected components (analogous to clusters), independent loops (analogous to filaments), and independent closed surfaces (representative of voids) which are topologically quantified by $\beta_0$, $\beta_1$, and $\beta_2$, respectively. Aforementioned figure indicates that the so-called connected components have more persistency rather than independent closed loops and  surfaces. In other words, a higher value of $\vartheta_{birth}$ corresponds to a higher value of $\vartheta_{pers}$, indicating that high-density regions are more persistent against the filtration process, similar to the temporal evolution of hierarchical clustering. In the upper right panel of  Fig.~\ref{fig:F2}, we plot the extracted $\beta_k$ from the fiducial realization as a function of filtration density $\delta + 1 = 10^\vartheta$ for better visualization\footnote{Throughout the analysis of Quijote field, we first map the $\delta$ to the $\log(\delta +1)$ and then apply the filtration.}. The blue solid line, orange dashed line and green dashed-dotted line indicate the $\beta_0$, $\beta_1$, and $\beta_2$, respectively. 
As the filtration threshold decreases, the number of independent topological features gradually increases until they reach their maximum value, after which they decrease as the proximity parameter approaches its minimum value.  The lower panel of Fig.~\ref{fig:F2} shows the $B_k$ and $P_k$ versus threshold, quantitatively confirming the representative behavior observed in the persistence diagram (upper left panel).

\subsection{Imprint of Massive Neutrinos on the PH vectorization} 
In the previous subsection, we discussed the topological properties of the fiducial model in the Quijote simulations and examined the overall behavior of $(\beta_k, B_k, P_k)$. Now, we will investigate the impact of massive neutrinos on the simulated density fields using our PH vectorization. Before proceeding, let's consider what we might expect from the effect of the summed neutrino mass on the topological invariants of the density field when performing super-level filtration. The large thermal velocities of massive neutrinos, combined with a decrease in the conventional dark matter density within the total mass content of the Universe, result in reduced clustering of the total matter field, especially at scales below their free-streaming length. This effect is further supported by the reduced depth of total matter potential wells in the presence of massive neutrinos. Consequently, we anticipate that the signature of massive neutrinos will manifest differently across all topological invariants for the $m$ and $cb$ fields. The presence of massive neutrino particles directly affects the $m$ field, which is a relevant observable field for surveys that measure the total matter field through weak lensing effects \citep{2008ARNPS..58...99H}. The analysis of the $cb$ field is also motivated by surveys that measure the clustering of halos or galaxies \citep{Kilo-DegreeSurvey:2023gfr}. Dark matter is typically traced by visible matter within gravitationally bound dark matter halos. However, galaxies do not always follow the associated mass distribution exactly. To account for this discrepancy, the mathematical description of this phenomenon often involves introducing halo bias and corresponding scale-dependent formalisms \citep{2021JCAP...03..022L,2018PhR...733....1D,2023MNRAS.524.1746L}. Evidently, halos trace the $cb$ field more closely than the $m$ field due to the large scale of massive neutrinos' free-streaming. Consequently, the topological and geometrical properties of the $m$ field are not always fully observable. Therefore, depending on the types of surveys and the scales used for evaluation, both $cb$ and $m$ fields become important for further examination.

 \begin{figure*}
 	\centering
 	\includegraphics[width=2.1\columnwidth]{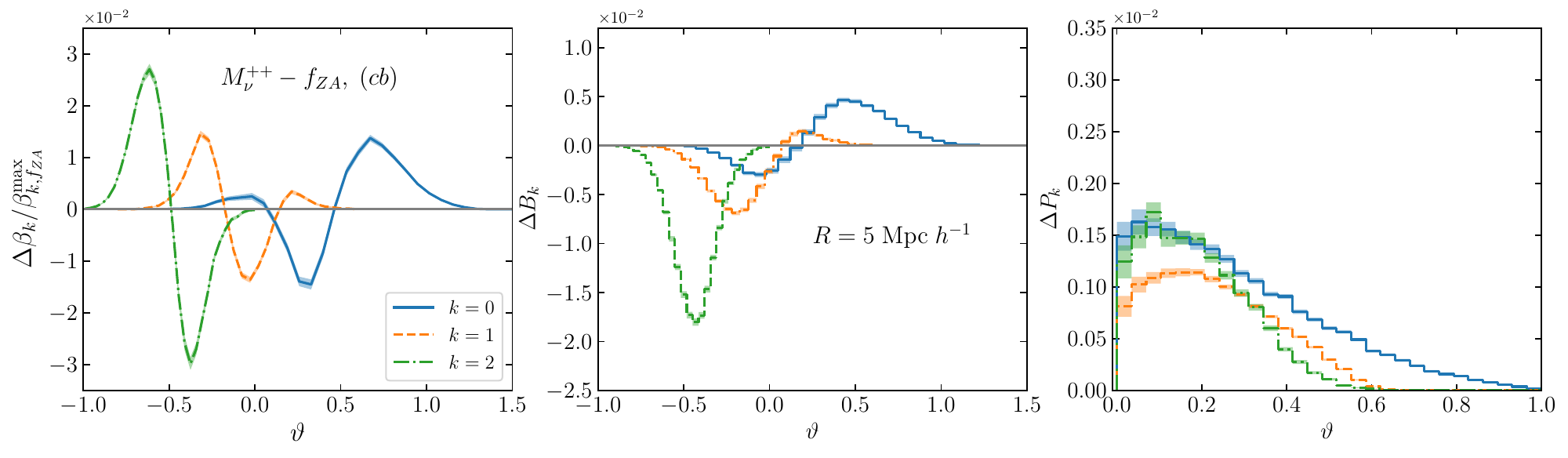}
 	\includegraphics[width=2.1\columnwidth]{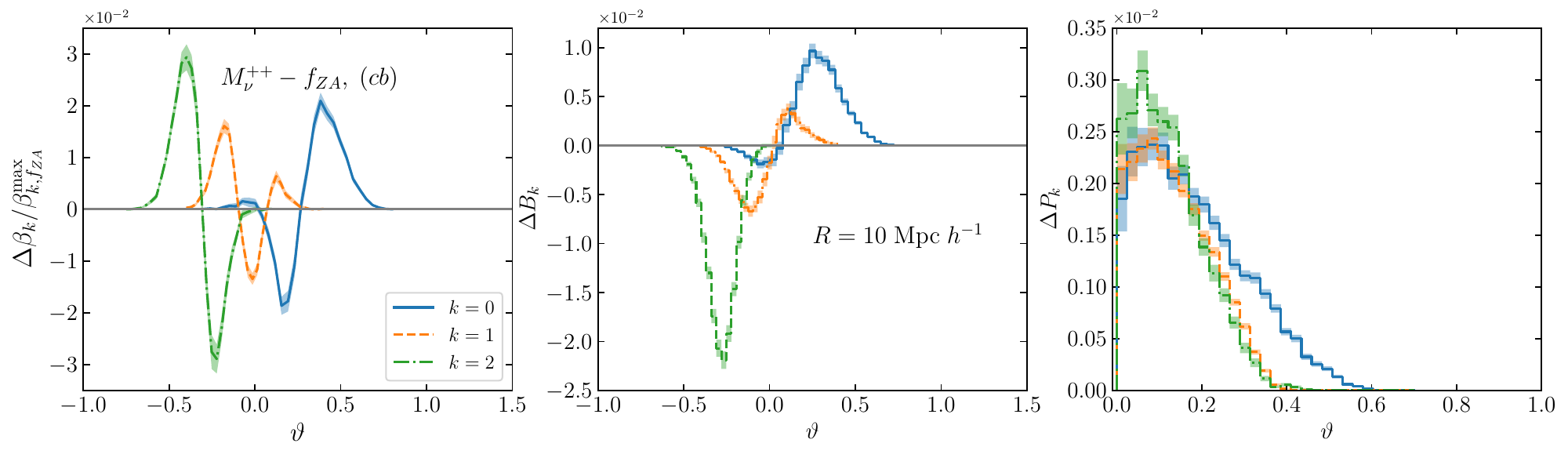}
 	\caption{Same as Fig. \ref{fig:F3_new}  just for the $cb$ part of simulated field.}
 	\label{fig:F3}
 \end{figure*}
In this subsection, we apply the statistics defined earlier, namely $(\beta_k,B_k,P_k)$, to both the $m$ and $cb$ fields in the presence of massive neutrino particles and compare these to the fiducial model to quantify the impact of neutrino mass on the simulated fields. We begin with computing the $\beta_k$, $B_k$, and $P_k$ measures over the $m$ field when the massive neutrino particles ($M_{\nu}^{++} = 0.2 \ \rm eV$) are included in the fiducial simulation. The relative differences for PH vectorization computed for $m$ field, compared to the fiducial simulation, are illustrated in Fig.~\ref{fig:F3_new}. For the better visualization, we divide the $\beta_k$ by $\beta_{k,f_{ZA}}^{\rm max}$, where the subscript ``$f_{ZA}$" denotes to the Zel'dovich approximation fiducial model \citep{Quijote_sims}. The upper and lower rows correspond to the smoothing scales, $R=5$ Mpc $h^{-1}$ and $R=10$ Mpc $h^{-1}$, respectively. The left panel of Fig.~\ref{fig:F3_new} illustrates the Betti-curves versus threshold.  For the smallest smoothing scale, the $\beta_2$ (2-hole illustrated by green dashed-dotted line) demonstrates the higher difference compared to other measures,  while the relative difference for independent closed surfaces is significantly suppressed by increasing the smoothing scale. Furthermore, the number of 2-holes is elevated to the higher thresholds.
The abundance of $\beta_0$ and $\beta_1$ remains relatively stable with changes in the smoothing scale, at least when transitioning from $R=5$ Mpc $h^{-1}$ to $R=10$ Mpc $h^{-1}$. Additionally, we present the $B_k$ and $P_k$ in the middle and right panels of Fig.~\ref{fig:F3_new}, respectively. These results confirm that the 2-hole statistics are particularly sensitive to the smoothing scale.
The imprint of the massive neutrinos from the perspective of $B_2$  is largely diminished with increasing smoothing scale for $\vartheta\le -0.5$  leading to a shift in the abundance of independent closed surfaces to higher thresholds. This behavior can be explained by the fact that in our super-level filtration pipeline, the spatial extension of 2-holes is considerably affected by the imposed smoothing scale for the negative density contrast compared to the 0-holes and even the 1-holes (see the lower panels of Fig. \ref{fig:F22}). We argue that the impact of massive neutrinos on the $m$ field manifests as a suppression of clustering due to the presence of massive neutrinos. Consequently, we expect a reduction in the number of $\beta_2$ compared to the fiducial cosmology (as indicated by the green dashed-dotted line in the left panels of Fig. \ref{fig:F3_new}). The contribution of massive neutrinos yields a decrease in the fraction of both under-dense and over-dense regions in the $m$ field. These findings are consistent with the MFs analysis conducted for Quijote simulations \citep{Liu2022ProbingMN}. 

The $cb$ density field, with massive neutrino particles included, can be analyzed in the simulation. Fig.~\ref{fig:F3} illustrates the relative difference in PH vectorization compared to the Zel'dovich approximation fiducial model. The maximum difference is observed for $\beta_2$ (2-hole), and it remains relatively stable across the smoothing process.  This result suggests that the topological properties of the so-called voids characterized by $\beta_2$, effectively trace the impact of neutrino mass on the LSS, which is consistent with the influence of massive neutrinos on the $cb$ field. The middle and right panels of Fig.~\ref{fig:F3} are devoted to $B_k$ and $P_k$, respectively. 
\begin{figure*}
	\includegraphics[width=2.1\columnwidth]{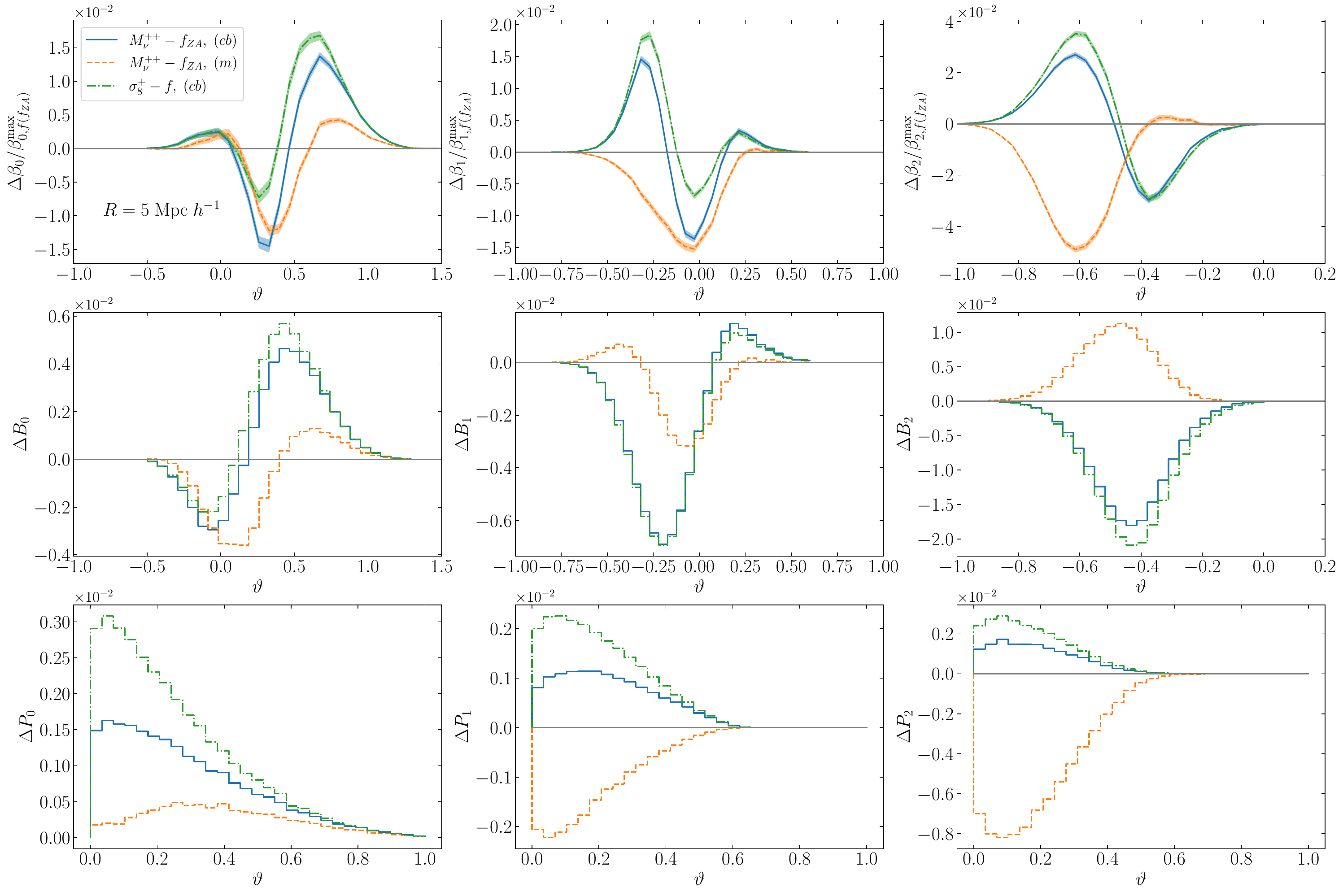}
	\caption{\label{fig:fig4} The upper rows indicate results for betti curves, while the middle rows are for $B_{k}$. The lower panels illustrate the $P_k$. The blue solid line corresponds to the difference between $cb$ part of $M^{++}_{\nu}$ simulation and fiducial cosmology. The orange dashed line is similar to the blue solid curve just for the total matter density field. The green dashed-dotted is associated with $\sigma_8^+$ simulation. Here we adopted $R=5$ Mpc $h^{-1}$.  }
\end{figure*}

To interpret our results, we should first examine the impact of massive neutrinos on the clustering of structures, particularly at scales smaller than $\lambda_{fs}$. As previously mentioned, the suppression of clustering at these scales due to massive neutrinos leads to a decrease in $\sigma_8$ for the total matter field. In the case of the Quijote simulations, the value of $\sigma_8$ for the total matter field ($m$) with massive neutrinos is kept similar to the fiducial cosmology value. This suggests that $\sigma_8$ for the CDM+baryons component is relatively higher.  Including massive neutrino particles in the corresponding pipeline for the Quijote suite is assumed to satisfy the fixed value for the $\sigma_8$ condition compared to the fiducial simulations. Additionally, the $\sigma_8$ value is adjusted by tuning the amplitude of primordial scalar perturbations \citep{2021JCAP...03..022L}. Therefore, for the $cb$ (CDM+baryons) field, denoted as $\sigma_{8}^{(cb)} = 0.846$, this value is higher compared to the fiducial value of $\sigma_8 = 0.834$ for the Quijote simulations with massive neutrinos. Subsequently, the presence of massive neutrino particles in the density field not only varies the population of the $k$-holes but also, changes the persistence pairs from topological viewpoints. As depicted in Fig. \ref{fig:F3}, a higher number of births are expected to appear at higher thresholds for 0-holes, while 2-holes mostly emerge at lower thresholds. We also expect to have a migration in the population of 2-holes from higher thresholds to the lower mass represented by the $\beta_2$ curve, whereas, this situation behaves in the opposite way for the connected component quantified by the $\beta_0$ diagram (the blue solid line) and the peak location of the 0-hole population moves toward the higher threshold regions. The enhancement on both tails and a reduction in the intermediate thresholds are realized for the $\beta_1$ curve, which analogously corresponds to the filaments (the orange dashed line in Fig. \ref{fig:F3}). By taking into account the inherent characteristics of $m$ and $cb$ fields, it can be seen that the impact of massive neutrinos on the PH vectorization, specifically for the $\beta_2$ at negative thresholds, is played in opposite directions. This behavior will also be confirmed by the Fisher forecasts analysis performed in the next section.   
In addition to the significant considerations regarding void-finding algorithms mentioned in section \ref{sec:imp}, the observed void tracers exhibit non-trivial behaviors affected by massive neutrinos \citep{2015JCAP...11..018M,2017MNRAS.465.4281P,2019MNRAS.488.4413K,pranav2019topology,Xu2019,Contarini2020CosmicVI}. The specific characteristics of simulations are expected to show significant variations in response to the inclusion of massive neutrinos in the LSS \citep{Liu2022ProbingMN}.

The influence of smoothing scale on the PH vectorization for $cb$ field is elucidated by comparing the upper and the lower rows in Fig. \ref{fig:F3}. The relative significance of PH properties associated with $\beta_2$, along with its robustness with respect to the smoothing procedure compared to other parts of the PH vectorization, indicates that 2-holes can be a qualified measure for evaluating massive neutrinos.
  \begin{figure*}
  	\includegraphics[width=2.1\columnwidth]{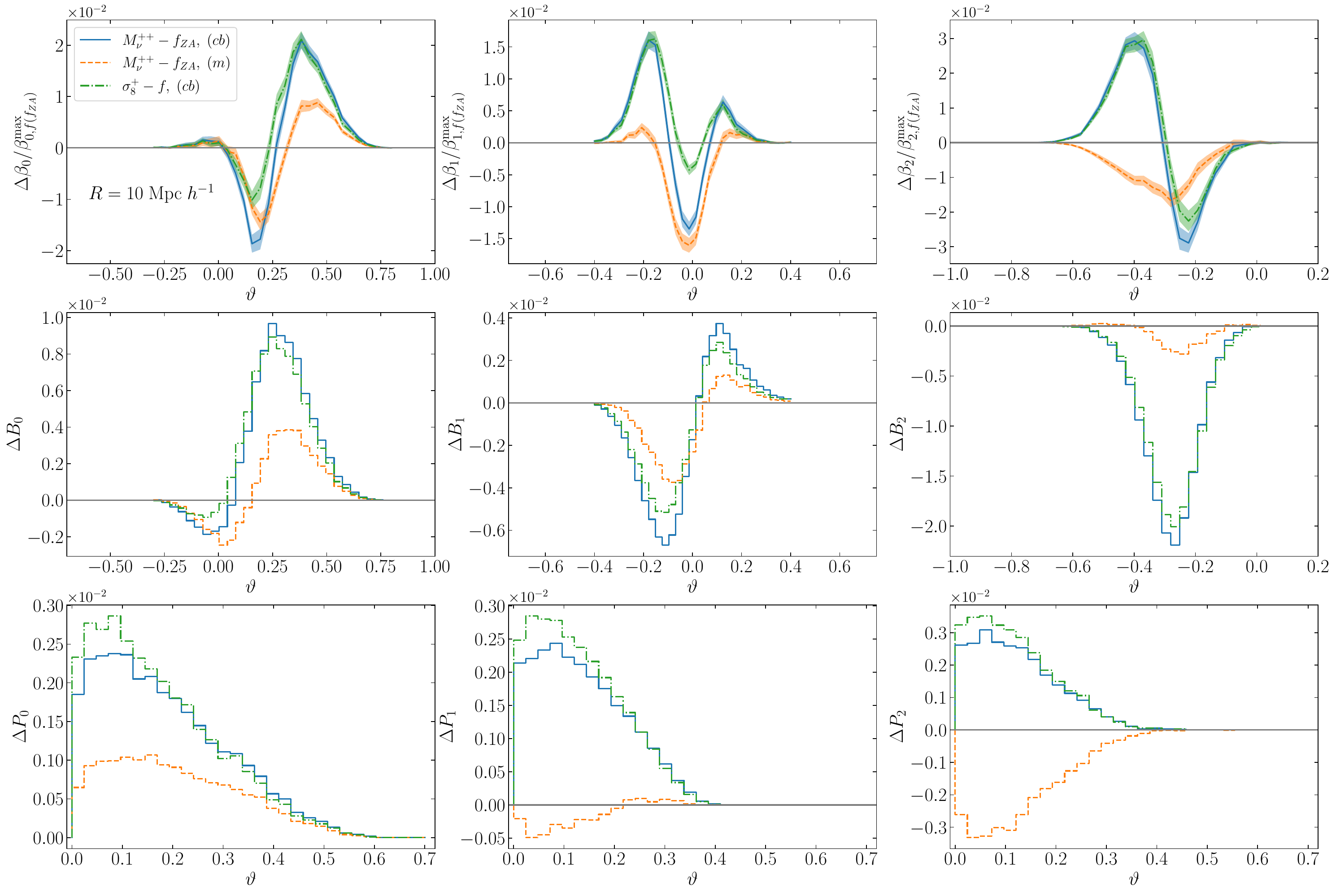}
  	\caption{\label{fig:fig42} Same as Fig. \ref{fig:fig4} just for $R=10$ Mpc $h^{-1}$.  }
  \end{figure*}

According to the functional form of the $\sigma_8$, it is well-known that this quantity would  possibly mimic the effect of massive neutrino particles leading to degeneracy in the plane of $(M_{\nu}, \sigma_8)$, especially when considering standard two-point clustering statistics \citep{2011MNRAS.418..346M,2014JCAP...02..049C}. Therefore, it is important to examine how the PH vectorization responds to such degeneracy. To address this issue, we use another subset of the Quijote simulations with a different $\sigma_8$ value, denoted by $\sigma_8^+ = 0.849$, but without massive neutrinos. Notably, $\sigma_8^+$ resembles $\sigma_8^{(cb)}$, which corresponds to the $cb$ field of the $M_{\nu}^{++}$ simulation. Motivated to assess the sensitivity of PH vectorization to $\sigma_8$ and the presence of $M_{\nu}$, we compare the behavior of $(\beta_k, B_k, P_k)$ for fields with massive neutrinos (i.e., $m$ and $cb$) against the case with $M_{\nu} = 0$ and $\sigma_8^+$. We aim to identify any promising signatures via the PH lens that could help break or at least reduce the degeneracy between $\sigma_8$ and $M_{\nu}$. 
 Fig. \ref{fig:fig4} indicates the $\beta_k$ (upper row), $B_k$ (middle row) and $P_k$ (lower row) versus $\vartheta$. The blue solid line corresponds to the difference between results for the PH measures applied to the $cb$ part of the field, including the massive neutrinos ($M^{++}_{\nu}$) and fiducial model, while the orange dashed line is associated with $m$ field. The green dashed-dotted line shows for field with $\sigma_8^+$ in the absence of the massive neutrinos.  The overall behavior of $(\beta_k,B_k,P_k)$ for $cb$ field for both  $\sigma_8^+$ and $M_{\nu}^{++}$ simulations are almost similar, and by increasing the value of $R$, mentioned difference becomes even lower (Figs. \ref {fig:fig4} and \ref{fig:fig42}).   
 The 1- and 2-holes statistics for $m$ field demonstrate distinguished behavior compared to the $\sigma_8^+$ simulations for $R=5$ Mpc $h^{-1}$. Increasing the smoothing scale yields a significant reduction in the  $\Delta \beta_2$, $\Delta B_2$ and $\Delta P_2$.

In order to give a quantitative measure for evaluating the signature of massive neutrinos based on the PH vectorization, we also integrate out the $\vartheta$-dependency through  the moment definition as: 
\begin{eqnarray}\label{eq:moment}
	\Xi_{k}^{(n)}\equiv \int d\vartheta \;\vartheta^n\; \mathcal{P}(\beta_k(\vartheta))
\end{eqnarray}
where $\mathcal{P}$ is the probability density function computed for $\beta_k$. 

\begin{figure}
	\includegraphics[width=1\columnwidth]{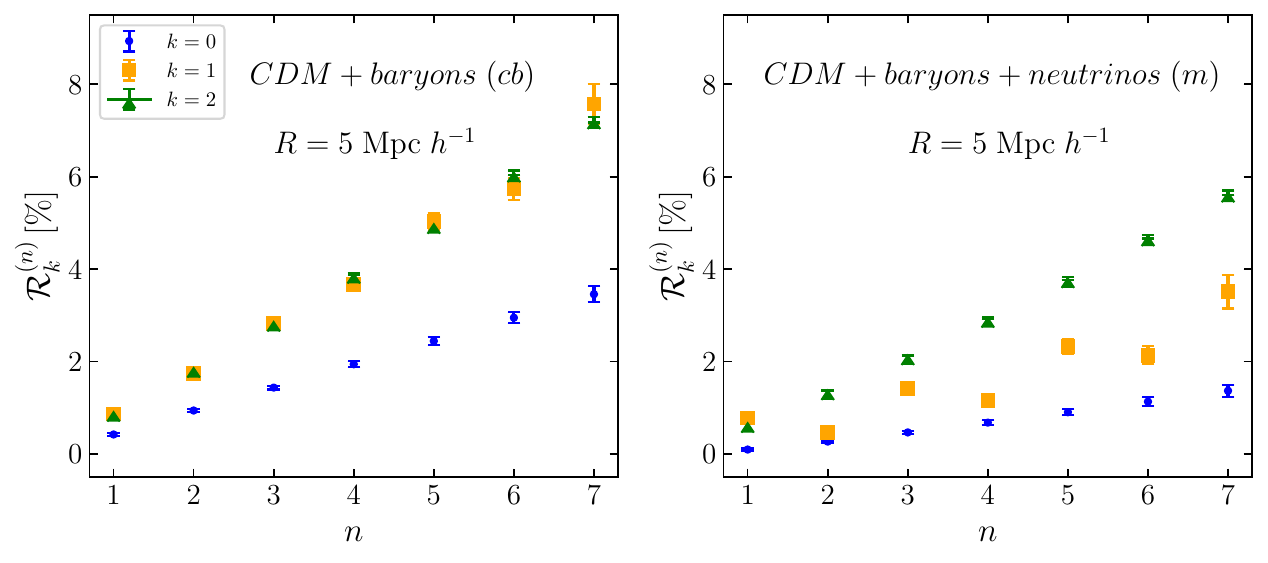}
		\includegraphics[width=1\columnwidth]{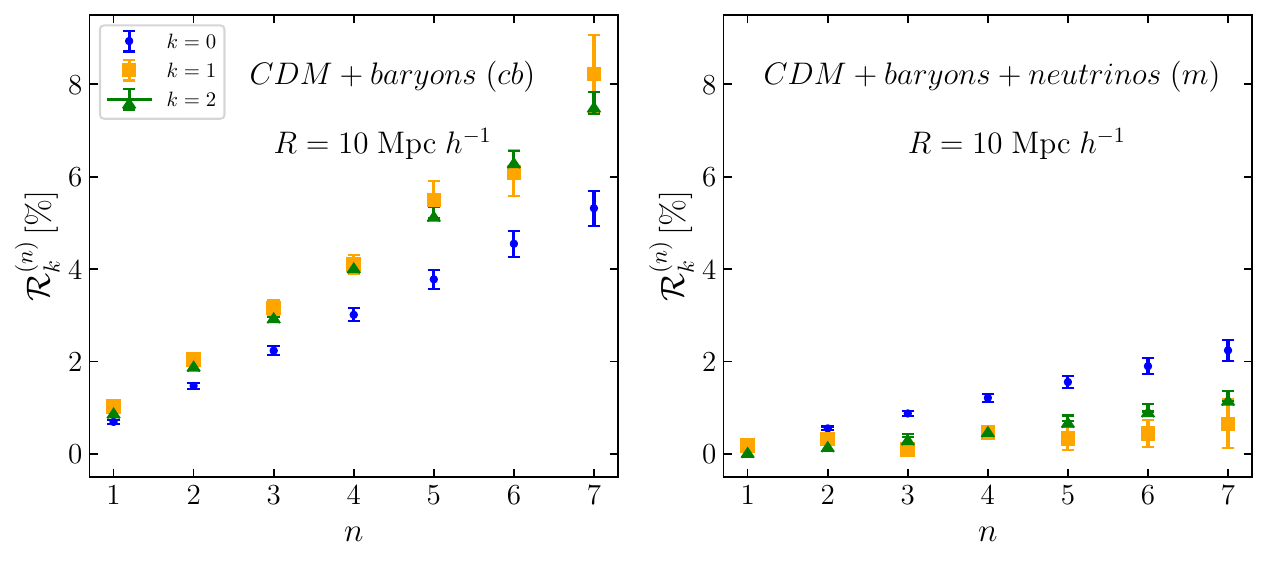}
	\caption{The percentage relative difference of Betti marginalized moments for $cb$ (left column) and $m$ (right column) fields of $M^+_{\nu}$ simulation with respect to the fiducial field. The blue filled circle, orange filled square, and green filled triangle symbols are for $\beta_0$, $\beta_1$, and $\beta_2$, respectively. The upper and lower rows correspond to $R=5$ Mpc $h^{-1}$ and $R=10$ Mpc $h^{-1}$, respectively. }\label{fig:F55}
\end{figure}

Generally, by using the moments or cumulants, we can encapsulate the statistical information encoded by probability distribution function. A motivation for us to define $\Xi_k^{(n)}$ in our study is to adopt a method that reduces the size of the feature vector, particularly for simulation-based inferences and Machine Learning approaches that will be part of our future research. From the perspective of statistical information and the cumulant expansion theorem \citep{matsubara2003statistics}, moments and connected moments are used to characterize the shape and tendency of associated probability density functions (PDF). Since we do not have the analytical form of $\mathcal{P}(\beta_k(\vartheta))$, we use the proper estimator to compute the various orders of moments. Subsequently, to reconstruct the analytical form of the PDF for further analysis from moment/cumulant expansion, computing moments is crucial. The robustness of higher-order moments necessarily requires a larger number of samples. From the error propagation routine in probability statistics, considering the significance level of moments and cumulants imposes an upper bound on the value of $n$ for a fixed size of data sets. Accordingly, we define $\mathcal{R}_k^{(n)}\equiv|\Xi_{k,M^+_{\nu}}^{(n)}-\Xi_{k,fid}^{(n)}|/(\Xi_{k,fid}^{(n)})$ and plot this quantity for the $m$ and the $cb$ fields of $M^+_{\nu}$ simulation in Fig. \ref{fig:F55}. The signature of massive neutrinos in the  2-holes are higher than the connected component captured by $\beta_0$ in the case of $m$ field for $R=5$ Mpc $h^{-1}$.  As an illustration for $n=5$, we obtain almost $\sim$4\% when the $\beta_2$ is adopted. Taking into account the $cb$ component implies the percentage relative differences for $5$th moment  are almost $\sim$5\% and $\sim$2.5\% when the $\beta_{(1,2)}$ and $\beta_0$ are considered, respectively for smallest smoothing scale at $z=0$. 

To elaborate the impact of smoothing scale on the $\mathcal{R}_k^{(n)}$, the upper and lower rows of Fig. \ref{fig:F55} are associated to $R=5$ Mpc $h^{-1}$ and $R=10$ Mpc $h^{-1}$, respectively. Applying a higher smoothing scale on the $m$ field decreases the difference between the fiducial cosmology and the $M^+_{\nu}$ simulation (see the right column of Fig. \ref{fig:F55}). As indicated in the left column of  Fig. \ref{fig:F55}, increasing the smoothing scale shows that the $\beta_k$'s persist for the $cb$ field, which is almost compatible with the results depicted in the left column of Fig. \ref{fig:F3}. 
It is important to note that the independent closed surface exhibits greater sensitivity to massive neutrinos, as indicated by $\mathcal{R}_k^{(n)}$ in the context of $\Xi_{k}^{(n)}$ measure for  $R=5$ Mpc $h^{-1}$. According to the right column of Fig. \ref{fig:F55},we can infer that for the $m$ field at  $R=5$ Mpc $h^{-1}$, $\beta_2$ demonstrates a strong response to the presence of massive neutrinos. In contrast, at $R = 10$ Mpc $h^{-1}$, the criterion indicates a milder level of deviation, with $\beta_0$ showing almost profound sensitivity. For the $cb$ field, the $R$-dependency of $\mathcal{R}_k^{(n)}$ is less pronounced (left column of Fig. \ref{fig:F55}).

\section{Quantifying information content: Fisher Forecasts}\label{sec:fish}
In this section,  relying on the Fisher matrix analysis, we quantify the capability of topological measures $(\beta_k,B_k,P_k)$ and their combinations on constraining the cosmological parameter denoted by a row vector as $\theta:\{\Omega_m,\Omega_b,\sigma_8,h,n_s,M_{\nu}\}$.  The element of the Fisher information matrix is defined by:
\begin{eqnarray}
	F_{ij} = -\left \langle \frac{\partial^2\ln\mathcal{L}}{\partial\theta_i \; \partial\theta_j}\right \rangle
\end{eqnarray}
where $\mathcal{L}$ is the Likelihood distribution. Assuming the multivariate normal distribution for the Likelihood function, we have:
\begin{eqnarray}
	F_{ij} = \frac{\partial\boldsymbol{\mathcal{A}}^T}{\partial\theta_i}\; C^{-1} \; \frac{\partial\boldsymbol{\mathcal{A}}}{\partial\theta_j}
\end{eqnarray}
where $\boldsymbol{\mathcal{A}}$ represents the data vector consisting of any desired selection and combination of PH vectorization $(\beta_k,B_k,P_k)$. Also $i$ and $j$ run from 1 to 6. The $C$ denotes  covariance matrix constructed from the data vector, and it's elements reads as $C_{pq}=\langle (\boldsymbol{\mathcal{A}}_p-\langle \boldsymbol{\mathcal{A}}_p\rangle)(\boldsymbol{\mathcal{A}}_q-\langle \boldsymbol{\mathcal{A}}_q\rangle)\rangle$.
The size of covariance matrix is directly related to the threshold binning ($N_{bins}$) which quantifies the bin resolution of  each topological components $(\beta_k,B_k,P_k)$ for a given $k$, and the number of  measures considered in the data vector. Consequently for our case, we have $p,q\in\{1,2,3,...,N_{tot}\}$. As an illustration, since the $k\in\{1,2,3\}$ for each element of $\boldsymbol{\mathcal{A}}$ and taking into account all combination of vector types, we achieve  $N_{tot}=(9\times N_{bins})$. To estimate the unbiased covariance matrix considering the $N_{sim}=5000$ realizations of the fiducial simulations in our pipeline, we use ${C}^{-1}\to\frac{N_{sim}-N_{tot}-2}{N_{sim}-1}C^{-1}$ \citep{2007A&A...464..399H}. To calculate the partial derivative of the vector $\boldsymbol{\mathcal{A}}$ with respect to the cosmological parameters, we use the following approximation:
\begin{eqnarray}
	\frac{\partial\boldsymbol{\mathcal{A}}}{\partial\theta_i}\; \simeq \frac{\boldsymbol{\mathcal{A}}(\theta_i^+)-\boldsymbol{\mathcal{A}}(\theta_i^-)}{\theta_i^+-\theta_i^-}
\end{eqnarray}
where $\boldsymbol{\mathcal{A}}(\theta_i^+)$   and $\boldsymbol{\mathcal{A}}(\theta_i^-)$ represent the extracted data vector from simulations in which the value of their $\theta_i$ parameter is higher and lower than the fiducial value, respectively. Also, to calculate the partial derivative with respect to $M_{\nu}$, we use:                                                                     
\begin{eqnarray}
	\frac{\partial\boldsymbol{\mathcal{A}}}{\partial M_{\nu}}\; \simeq \frac{\boldsymbol{\mathcal{A}}(M_{\nu}^+)-\boldsymbol{\mathcal{A}}(\theta_{ZA})}{0.1}
\end{eqnarray}
where $\boldsymbol{\mathcal{A}}(M_{\nu}^+)$ indicates the obtained data vector from massive neutrinos simulation with $M_{\nu} = 0.1 $ eV, and  $\boldsymbol{\mathcal{A}}(\theta_{ZA})$ represent the extracted data vectors from fiducial simulations which their   initial conditions are generated with Zel'dovich approximation. To estimate the partial derivatives, for each of the model parameters, we have used 500 corresponding realizations (for more details see \cite{Quijote_sims,2020MNRAS.495.4006U}). 

\begin{figure}
	\includegraphics[width=0.8\linewidth]{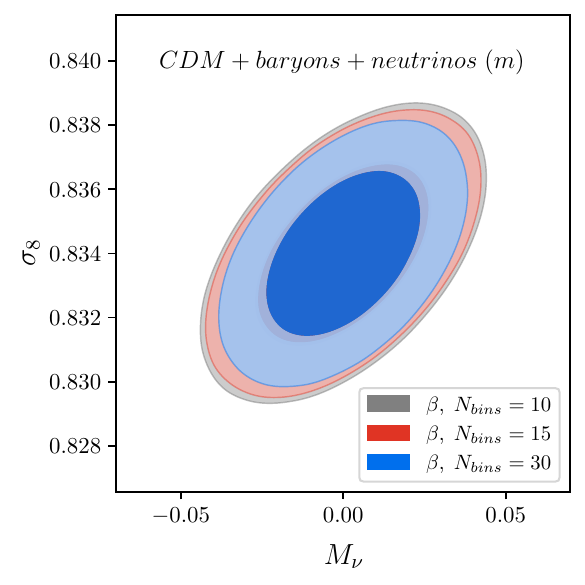}
		\caption{The influence of $N_{bins}$ on the $(M_{\nu},\sigma_8)$ plane in the Fisher forecast for $m$ field. Here we took $R=5$ Mpc $h^{-1}$ and $\boldsymbol{\mathcal{A}}:\{\beta_0,\beta_1,\beta_2\}$.}
	\label{fig:F9}
\end{figure}

Table \ref{ta:Table0} reports the range of thresholds used for the Fisher forecasts analysis. Each interval has been achieved by considering the removing rare events appeared in the both tails of ranges in the deep positive and negative thresholds. Since, by definition, the $B$ and $P$ are accumulative statistics and for the first bin interval, whose values equate to unity,  we neglect the first bin value to prevent the singularity in the computed Fisher matrix. While for the $\beta$ measure such discrepancy does not occur. We divide the threshold ranges reported in Table \ref{ta:Table0} with the $N_{bins}=\{10,15,30\}$, to examine the influence of bin size of $\boldsymbol{\mathcal{A}}$ on the row vector. For the $N_{bins}=15$ and for each bin, we have at least $\mathcal{O}(1000)$ topological features to ensure the reliability of having a Gaussian distribution\footnote{We have verified the significance level of Gaussianity using the Kolmogorov-Smirnov test and found no significant deviation from the Gaussian hypothesis at the $95\%$ CL.}. The Fisher forecast results generally depend on the number of bins. To estimate reliable constraints on the free parameters, the optimal value of $N_{bins}$ should be determined under the assumption of Gaussianity. In this analysis, we naively examine the influence of $N_{bins}$ on the Fisher forecast results by directly computing the confidence intervals for the relevant free parameters. We begin our investigation by constraining the pair parameters $M_{\nu}$ and $\sigma_8$. Our analysis considers a redshift of $z = 0$ and a smoothing scale of $R = 5$ Mpc $h^{-1}$. Fig. \ref{fig:F9} indicates the marginalized contours concerning the joint analysis of $\beta$'s in the plane of $(M_{\nu},\sigma_8)$ for different bin values. Our results demonstrate that, for the selected range of $N_{bins}$ in our analysis, the computed constraint regions are almost unaffected by changes in $N_{bins}$ up to the level of statistical uncertainty. A comprehensive evaluation of the influence of selecting $N_{bins}$ for supporting the robustness of quantifying the imprint of massive neutrinos will be left for future research.
\begin{table*}
	\centering
	\begin{tabular}{|c|c|c|c|c|c|c|c|c|c|}
		\hline
		$R$/ Threshold & $\vartheta_{\beta_0}$ & $\vartheta_{\beta_1}$ & $\vartheta_{\beta_2}$ & $\vartheta_{B_0}$ & $\vartheta_{B_1}$ & $\vartheta_{B_2}$&$\vartheta_{P_0}$&$\vartheta_{P_1}$&$\vartheta_{P_2}$ \\
		\hline
		$5$ Mpc $h^{-1}$ & [-0.2,1.0] &  [-0.5,0.3] &  [-0.8,-0.2] &  [-0.2,0.8] & [-0.4,0.4] &  [-0.6,-0.1]& [0.0,0.9]&  [0.0,0.5]&  [0.0,0.5] \\
		\hline
		$10$ Mpc $h^{-1}$ & [0.0,0.5] &  [-0.2,0.2] &  [-0.45,-0.1] &  [0.0,0.5] & [-0.2,0.2] &  [-0.45,-0.1]& [0.0,0.5]&  [0.0,0.35]&  [0.0,0.35] \\
		\hline
	\end{tabular}
	\caption{The domain of thresholds for various PH vectorization used in the Fisher forecasts for different smoothing scales.}\
	\label{ta:Table0}
\end{table*}

\begin{figure}
	\includegraphics[width=0.8\linewidth]{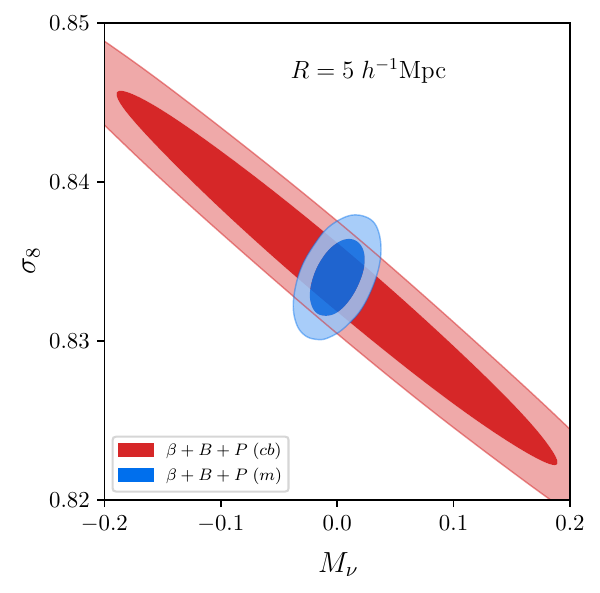}
	\includegraphics[width=0.8\linewidth]{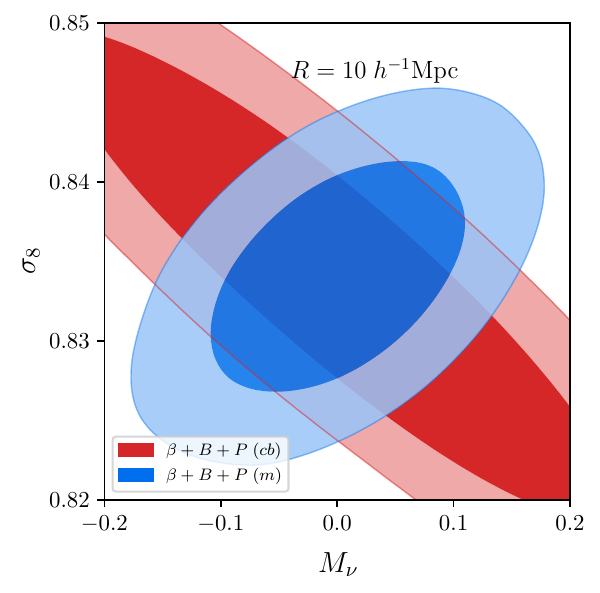}
	\caption{Upper panel is the Fisher forecast based on the joint analysis of $\beta+B+P$ for the $cb$ and $m$ fields at $1\sigma$ and $2\sigma$ confidence intervals for $R=5$ Mpc $h^{-1}$. The lower panel is the same as above just for $R=10$ Mpc $h^{-1}$. Here we adopted $N_{bins}=15$.}
	\label{fig:F77}
\end{figure}

In Fig.~\ref{fig:F77}, we compare the obtained constrains from joint analysis, $\beta + P + B$, on $(M_{\nu}, \sigma_8)$ parameters for $cb$ and $m$ density fields for two different smoothing scales. The results show that the constraints are tighter in the $m$ field compared to the $cb$ field, which reduces the degeneracy between $M_{\nu}$ and $\sigma_8$. Additionally, increasing the smoothing scale decreases the significance of the constraint on the desired parameter, as illustrated in the lower panel of Fig. \ref{fig:F77}. 

\begin{figure*}
	\includegraphics[width=2.1\columnwidth]{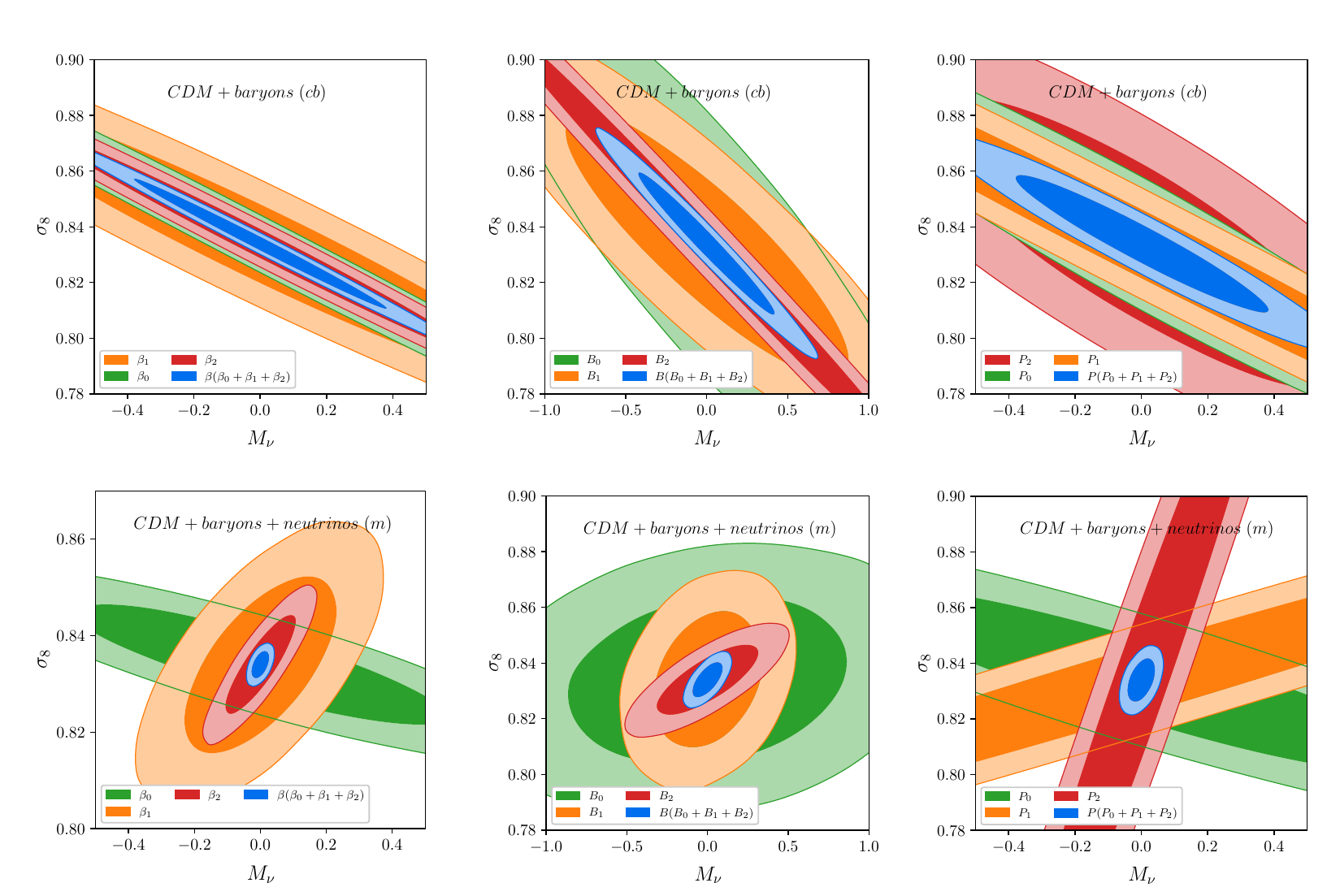}
	\caption{\label{fig:F6} The 68\% and 95\% confidence contours for $(M_{\nu}, \sigma_{8})$ obtained from various measures under the PH vectorization for both $cb$ and $m$ fields at redshift $z=0$ by the Fisher information analysis. The upper row depict constraints for $cb$ field, while the bottom row are devoted to  $m$ field constraints. Here we adopted $N_{bins}=15$ and $R=5$ Mpc $h^{-1}$.}
\end{figure*}

Fig.~\ref{fig:F6} presents the marginalized 68\% and 95\% confidence contours for the parameters $M_{\nu}$ and $\sigma_8$ for both $cb$ and $m$ fields. The upper row illustrates the constraints for the $cb$ density field, while the lower row shows the constraints for the $m$  density field. In the upper left column, we consider the various components of Betti curves including $\beta_0$, $\beta_1$, and $\beta_2$, and the associated joint analysis as the observable vector. The middle and right columns are the same as the left column, but for the $B_k$ and $P_k$ vectorization, respectively. 
 In this figure, for $cb$ field the directions of degeneracy between $M_{\nu}$ and $\sigma_8$ constraints obtained from various components of PH measures approximately are similar. This degeneracy also persists in the constraints resulting from their combinations.
  However, for the $m$ density field, the directions of degeneracy differ, leading to a reduction in the degeneracy between $M_{\nu}$ and $\sigma_8$ for the compound constraints. 

\begin{figure*}
	\includegraphics[width=1.3\columnwidth]{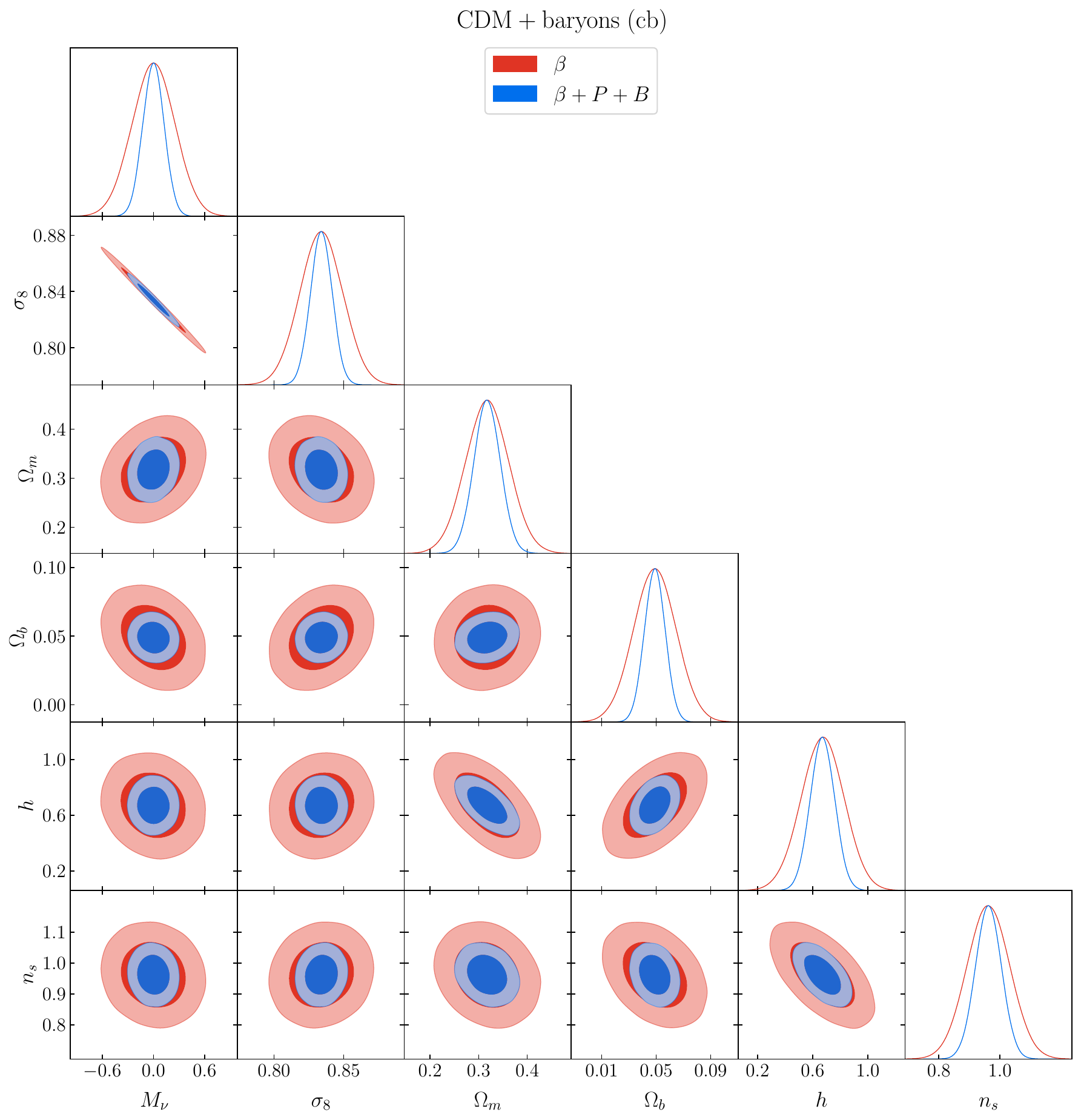}
	\includegraphics[width=1.3\columnwidth]{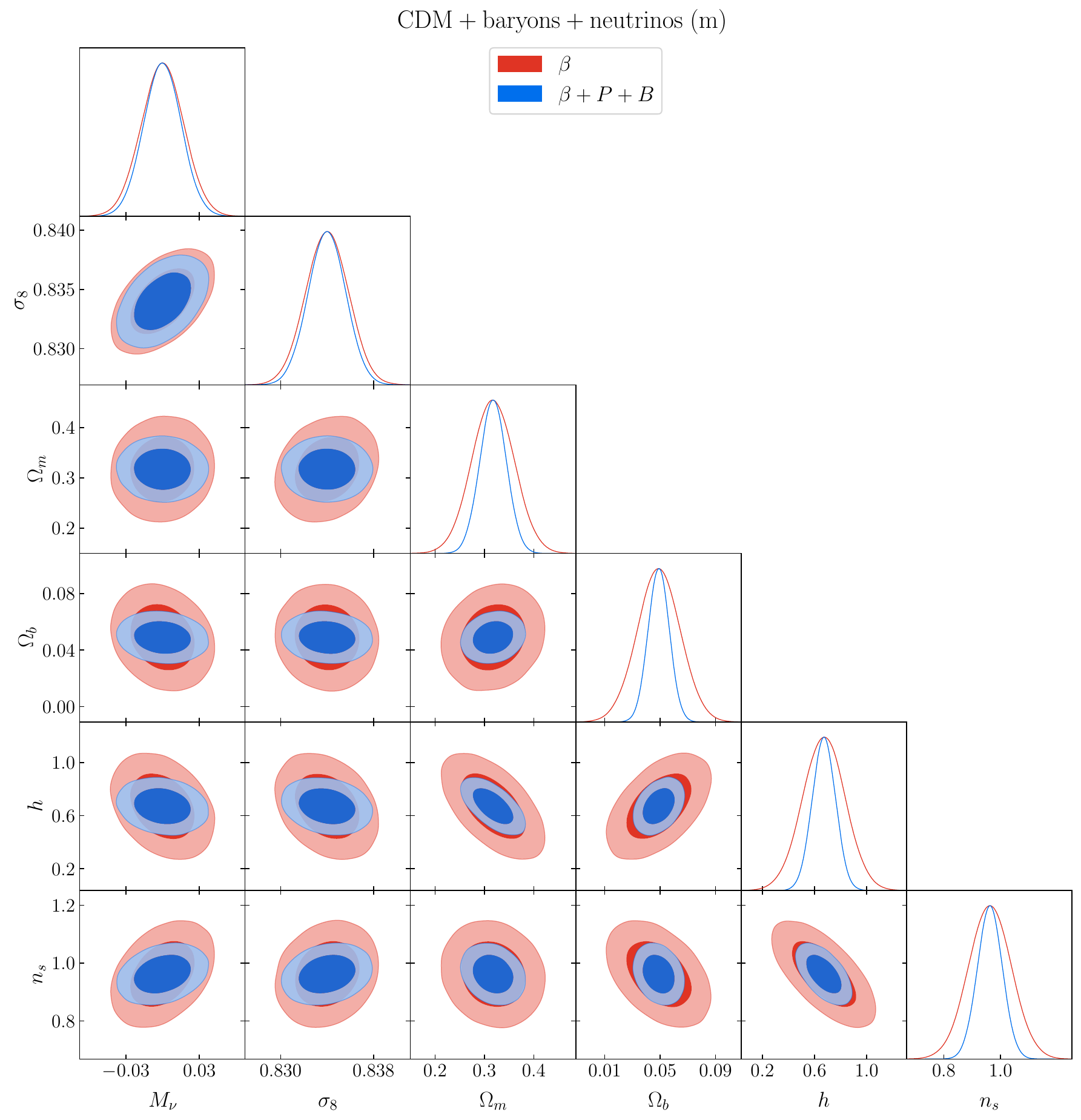}
	\caption{Fisher forecasts for cosmological parameters. The upper panel shows the confidence intervals on the cosmological parameters when the $cb$ field is considered. The lower panel is devoted to the total matter density field.  Here we adopted $N_{bins}=15$ and $R=5$ Mpc $h^{-1}$.}\label{fig:F88}
\end{figure*}

To enhance our analysis, we present the 2D constraints (at 68\% and 95\% CL) on the most relevant $\nu\Lambda$CDM parameters $(M_{\nu}, \sigma_8, \Omega_m, \Omega_b, h, n_{s})$. These constraints come from Fisher analysis of the $\beta$ parameter alone and the combined probes $\beta + B + P$ for the $cb$ field, as presented in the upper panel of Fig.~\ref{fig:F88}. The same results are depicted in  the lower panel of Fig.~\ref{fig:F88} for the $m$ field. The joint analysis, $\beta + P + B$, provides nearly twice the improvement in constraining cosmological parameters compared to using $\beta$ alone for the $cb$ field. We have also summarized the 68\% marginalized errors on the cosmological parameters from various probes for $R=5$ Mpc $h^{-1}$ in Table~\ref{ta:Table1} 
for better comparison, quantitatively. When assessing the ability of PH measures to constrain massive neutrinos relative to Minkowski Functionals, we find a notable improvement in the $cb$ field. However, the uncertainty in the $m$ field remains nearly identical to that obtained using Minkowski Functionals \citep{Liu2022ProbingMN}. On the other hand, MFs and PH (in the context of cubical complexes with super-level filtration) are the morphological tools that somehow quantify the morphological information of the excursion sets, and basically, carry out information regarding the N-point correlation functions.  
Thus, we do not anticipate that the constraints obtained from these morphological tools will differ dramatically. However, PH explores different aspects of the cosmic web compared to the MFs. This could potentially resolve some degeneracies and lead to tighter constraints when combined with other statistical tools, such as MFs and the power spectrum.

We have presented the marginalized 68\% errors on the cosmological parameters for smoothing scale $R=10$ Mpc $h^{-1}$ in Table~\ref{ta:Table2}. This allows us to evaluate the impact of increasing the smoothing scale on the constraining power of PH vectorization. The estimated errors for the parameters $M_{\nu}$, $\sigma_8$, and $n_{s}$ increase with $R=10$ Mpc $h^{-1}$, irrespective of the statistics used, as compared to the case with $R=5$ Mpc $h^{-1}$. For the other parameters, changes are less pronounced, and in some cases, a slight improvement in the errors is observed.

It should be noted that by using smoothing scales of $R=5$ and $10$ Mpc $h^{-1}$, we also account for information related to the non-linear regime and, consequently, non-Gaussian effects. Therefore, in this case, the Fisher matrix, which is based on the Gaussian hypothesis, does not provide an accurate estimation of the parameter errors. Therefore, for more accurate estimation, we need simulation-based inference (free likelihood) methods \citep{tejero2020sbi,2016arXiv160506376P,2019MNRAS.488.4440A,2020PNAS..11730055C}.

\begin{figure*}
	\includegraphics[width=0.7\columnwidth]{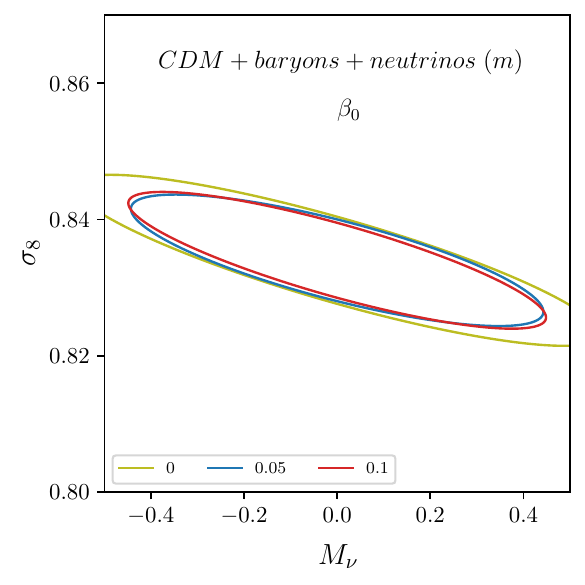}
	\includegraphics[width=0.7\columnwidth]{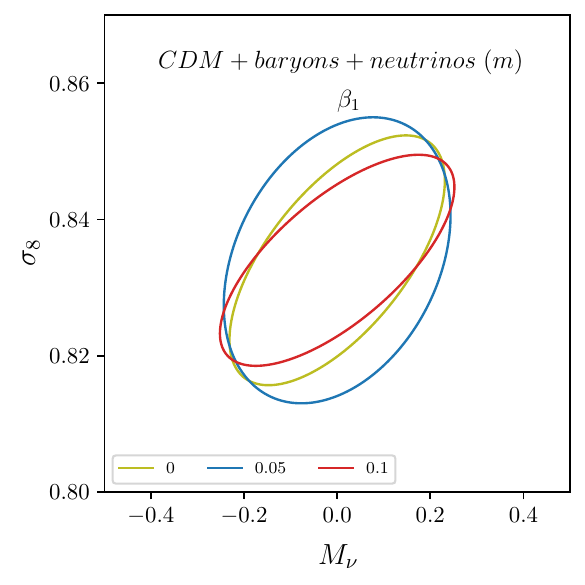}
		\includegraphics[width=0.7\columnwidth]{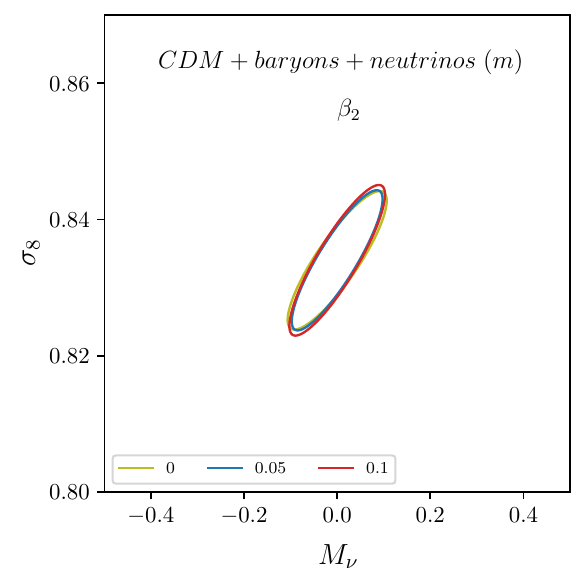}
			\includegraphics[width=0.7\columnwidth]{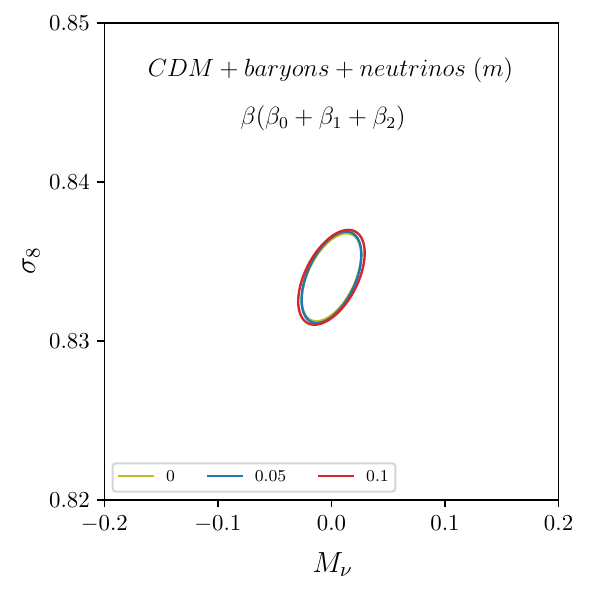}
	\caption{The Fisher forecast in the plane of $\sigma_8$ and $M_{\nu}$ for different pairs persistency levels of exclusion.  Here we adopted $N_{bins}=15$ and $R=5$ Mpc $h^{-1}$.}\label{fig:out}
\end{figure*}

To encapsulate the topological properties by applying the filtration procedure, the influence of outliers, such as noise, can affect the accurate estimation of topological features. To overcome the mentioned challenge, the filtration through the coarse-graining scales is modified by employing proper definition of the birth and the death thresholds \citep{edelsbrunner2000proceedings,edelsbrunner1994three,2020PhRvD.102h3537Z,wasserman2018topological}. Another approach to address the impact of noise is to remove persistent pairs whose persistency is below a given threshold. Here, we have performed the Fisher forecast analysis for various values of persistency thresholds. Taking into account the persistent pairs whose persistencies are $\vartheta_{pers}\ge0.0$ (our reference results), $\vartheta_{pers}\ge0.05$ (the $\sim65\%$ of pairs have been excluded) and $\vartheta_{pers}\ge0.1$ (the $\sim80\%$ of pairs have been excluded), we computed the uncertainty contours of the model's free parameters in the joint analysis, we computed the uncertainty contours of the model's free parameters in the joint analysis. Fig. \ref{fig:out} indicates the marginalized $68\%$ confidence contours on the $(M_{\nu},\sigma_8)$ plane obtained from our Fisher matrix calculations for the $m$ density field taking into account different values of minimum thresholds (the green, blue, and red lines correspond to the minimum threshold of 0, 0.05, and 0.1, respectively). Here, we used only Betti curves as data vectors in the Fisher analysis. It can be observed that varying the minimum persistency thresholds does not significantly alter the constraint regions. Therefore, the statistical information about parameter changes is primarily contained in pairs with high persistency. Consequently, in practical applications, low-persistence features that are susceptible to noise can be safely discarded without compromising sensitivity. We found that eliminating 80\% of the pairings results in a maximum relative difference in the constraining level, considering the joint analysis for $M_{\nu}$, of approximately 12\%. The associated uncertainty interval increases to 0.019 eV.

\begin{table*}
	\centering
	\begin{tabular}{|c|c|c|c|c|c|c|}
		\hline
		Statistics (field) & $M_{\nu} ({\rm eV})$ & $\sigma_8$ & $\Omega_m$ & $\Omega_b$ & $h$ & $n_{s}$ \\
		\hline
		$\beta$ ($cb$) & 0.2504 & 0.0153 & 0.0442 & 0.0156 &0.1546 & 0.0699 \\
		\hline
		$\beta$ ($m$) & 0.0172 & 0.0018 & 0.0427 & 0.0152 & 0.1617 & 0.0747 \\
		\hline
		$P$ ($cb$) & 0.2511 & 0.0162 & 0.0559 & 0.0163 & 0.1640 & 0.1128 \\
		\hline
		$P$ ($m$) & 0.0269 & 0.005 & 0.0564 & 0.0163 & 0.1650 & 0.1224 \\
		\hline
		$B$ ($cb$) & 0.2779 & 0.0168 & 0.062 & 0.0163 & 0.1380 & 0.1694 \\
		\hline
		$B$ ($m$) & 0.0610 & 0.0041 & 0.0615 &  0.0139 & 0.1709 & 0.1473 \\
		\hline
		$\beta + P + B$ ($cb$) & 0.1242 & 0.0077 & 0.027 & 0.0075 & 0.0878 & 0.0423 \\
		\hline
		$\beta + P + B$ ($m$) & 0.0152 & 0.0015 & 0.0267 & 0.0075 & 0.0886 & 0.0436 \\
		\hline
		
	\end{tabular}
	\caption{ The uncertainty interval computed by marginalization in the Fisher forecasts for the cosmological parameters  at 68\% level of confidence. Here we adopted $N_{bins}=15$ and $R=5$ Mpc $h^{-1}$.}
	\label{ta:Table1}
\end{table*}

\begin{table*}
	\centering
	\begin{tabular}{|c|c|c|c|c|c|c|}
		\hline
		Statistics (field) & $M_{\nu} ({\rm eV})$ & $\sigma_8$ & $\Omega_m$ & $\Omega_b$ & $h$ & $n_{s}$ \\
		\hline
		$\beta$ ($cb$) & 0.3520 & 0.0205 & 0.0411 & 0.0111 & 0.1370 & 0.1460 \\
		\hline
		$\beta$ ($m$) & 0.0927 & 0.0075 & 0.0386 & 0.0111 & 0.1275 &  0.1470 \\
		\hline
		$P$ ($cb$) & 0.3439 & 0.0223 & 0.0575 & 0.0167 & 0.1928 & 0.2048 \\
		\hline
		$P$ ($m$) & 0.1197 & 0.0138 & 0.0566 & 0.0164 & 0.1823 & 0.2038 \\
		\hline
		$B$ ($cb$) & 0.3139 & 0.0209 & 0.0450 & 0.0139 & 0.1611 &  0.1605 \\
		\hline
		$B$ ($m$) & 0.1970 & 0.0143 & 0.0448 &  0.0138 & 0.1604 & 0.1663 \\
		\hline
		$\beta + P + B$ ($cb$) & 0.1570 & 0.0100 & 0.0255 &  0.0066 & 0.0840 & 0.0926 \\
		\hline
		$\beta + P + B$ ($m$) & 0.0720 & 0.0047 & 0.0255 & 0.0066 & 0.0843 & 0.0937 \\
		\hline
		
	\end{tabular}
	\caption{The uncertainty interval computed by marginalization in the Fisher forecasts for the cosmological parameters  at 68\% level of confidence. Here we adopted $N_{bins}=15$ and $R=10$ Mpc $h^{-1}$.}
	\label{ta:Table2}
\end{table*}

\section{Summary and Conclusions}\label{sec:con}

Motivated by the various theoretical approaches and observational pipelines for evaluating the neutrino oscillation and associated masses, we relied on the computational topology to assess the massive neutrinos in the LSS. Recognizing the dominant impact of massive neutrinos on small-scale structure formation, we employed Betti numbers as a set of topological invariants, along with complementary quantitative measures proposed by \cite{biagetti2021persistence},  to analyze N-body simulations of the Quijote suite for massive neutrinos cosmology at $z=0$.  For different cosmological simulated fields, namely total mass density field ($m$), the CDM+baryons part ($cb$) in the presence of $M_{\nu}$, we have computed $(\beta_k, B_k,P_k)$. Comparing this PH vectorization of massive neutrinos
simulation and the fiducial model revealed that the independent closed surfaces (analogous to  voids) and corresponding complementary measures ($B_2$ and $P_2$) are more sensitive tracers to probe the footprint of neutrino mass scale in the LSS when the topological measures are used for smallest smoothing scale used in this research ($R=5$ Mpc $h^{-1}$) (Figs. \ref{fig:F3_new} and \ref{fig:F3}).

The modification of underlying simulated fields via smoothing scale has revealed that the abundances of $0-$ and $1-$holes for our super-level filtration pipeline remain relatively stable, while $\beta_2$ is significantly affected by different values of $R$. The spatial extension of the excursion sets for $\beta$'s is highly dependent to the threshold and for the lower values of $\vartheta$, the $\beta_2$ for $m$ field is more sensitive to $R$ as compared as the $\beta_0$ and $\beta_1$ (Fig. \ref{fig:F3_new} and the lower panels of Fig. \ref{fig:F22}).  

The degeneracy between $\sigma_8$ and $M_{\nu}$ complicates the use of two-point statistics of LSS for analyzing massive neutrinos. To assess whether the PH vectorization presented in this paper can address or at least mitigate this discrepancy, we have utilized two simulations $\sigma^+_8$ and $M^{++}_{\nu}$.  According to the results presented in Fig. \ref{fig:fig4} for $R=5$ Mpc $h^{-1}$,  the PH measures for these two simulations show similar behaviors for the $cb$ field. However, different trends are observed for the $m$ field. 
A promising result is that the topological features for $k=1$ and $k=2$ provide more reliable criteria to mitigate the influence of the mentioned degeneracy (Fig. \ref{fig:fig4}). Incorporating the impact of smoothing scale also revealed that eventually the $2$-holes and associated quantifications  in this study are suitable measures for evaluating the degeneracy related to the footprint of massive neutrinos on the LSS (right panels of Figs. \ref{fig:fig4} and \ref{fig:fig42}).


Marginalization over the  thresholds contribution in the Betti-curves (Eq. (\ref{eq:moment})) and based on definition of relative difference, $\mathcal{R}_k^{(n)}$, also confirmed that the higher ranks of Betti numbers are proper candidates for probing the massive neutrinos  for $R=5$ Mpc $h^{-1}$ (Fig. \ref{fig:F55}).  More precisely, for the fifth moment, the relative differences in $\beta_2$ are approximately $\beta_2$ are  almost  $\sim 4\%$ and $\sim5\%$ for $m$ and $cb$ fields, respectively  (Fig. \ref{fig:F55}). The higher value of $R$ results in suppressing the sensitivity of $\mathcal{R}_k^{(n)}$ to the signature of $M_{\nu}$ at zero redshift. Depending on the value of smoothing scale, different components of the PH vectorization can serve as effective indicators of the massive neutrino in LSS at $z=0$.  

We extended our analysis by computing the Fisher information matrix to  assess the capability of topological measures to constrain the relevant cosmological parameters. We divided the threshold intervals for each component of PH vectorization to the $N_{bins}$, ensuring that at least $\mathcal{O}(1000)$ topological features are captured (Table \ref{ta:Table0} for $R=5$ Mpc $h^{-1}$ and Table \ref{ta:Table1} for $R=10$ Mpc $h^{-1}$).  Based on the unbiased estimator of covariance matrix and doing the ensemble averaging on $N_{sim}=5000$ realizations, we examined the potential of PH vectorization to provide more precise constraints for different smoothing scales at $z=0$. To perform a preliminary assessment of $N_{bins}$ in PH vectorization tool, we considered three bin numbers for thresholds and computed the corresponding Fisher forecasts based on the joint analysis of Betti numbers. The results indicate that our findings are largely unaffected (less than $\sim 0.2\%$ for $\sigma_8$ and less than $\sim 1\%$ for $M_{\nu}$ to triple the amount of $N_{bins}$ (Fig. \ref{fig:F9})) at $z=0$ for $m$ field. Our analysis of $R$-dependency has shown that increasing the smoothing scale results in a decrease in the significance of constraints on the cosmological parameters in the $(M_{\nu},\sigma_8)$ (Fig. \ref{fig:F77}).

Betti curves, as one of the components of PH vectorization, have demonstrated better constraining power on the parameters compared to other measures. The combinations improve the constraints approximately twofold for the $cb$ field. However, using the $cb$ field, the degeneracy in the $(M_{\nu}, \sigma_8)$ plane persists. In this case, it is suggested to add new information, such as studying more than one map to capture the direct evolutionary footprint of the field. No similar degeneracy is observed for the $m$ field (Figs.  \ref{fig:F6}, \ref{fig:F88} and Table \ref{ta:Table1}).  For tracking the impact of noise in the filtration procedure, we have performed the Fisher forecasts for different values of persistency threshold exclusion. The highest relative difference in the uncertainty level for $M_{\nu}$, is around 12\% revealing the associated uncertainty interval increases to 0.019 eV (Fig. \ref{fig:out}).

Finally, our study demonstrates the potential of persistent homology as a valuable tool for analyzing complex cosmological datasets and extracting information for the impact of massive neutrinos on the LSS. This approach shows a significant improvement in constraining neutrino mass along with the other cosmological parameters. Due to the high capability of the PH approach to reveal the impact across a variety of scales in a field, it is worth considering the application of this method to current and upcoming high-precision cosmological surveys to explore different scenarios, even beyond the active three-flavor scheme.

We should point out that to set up a pipeline to compare the results with observations, the following tasks should be carried out \citep{2018PhR...733....1D}:\\
	1) Constructing the density field based on a robust algorithm.‌ \\ 
	2) A so-called halo finder to identify the matter halo should be implemented on the constructed density field in simulation. The simulated field includes the CDM+baryons ($cb$) in the presence of the massive neutrinos. Since the scales used to construct halos are much smaller than the free-streaming scale of massive neutrinos, the $cb$ field provides a reasonable contribution in the presence of massive neutrinos and is considered a suitable tracer for observations. In other words, at the scales we focus on, massive neutrinos have an implicit impact, and depending on the observable quantities considered, both the $cb$ and $m$ fields offer distinct advantages for separate investigation   \citep{2020JCAP...03..040H,2021JCAP...01..009P,2012PhRvD..85f3521I,2014JCAP...03..011V,2014JCAP...02..049C,2018JCAP...09..001V},  \\
	3) A halo occupation distribution and sub-halo abundance matching approaches must be taken into account to build mock galaxy catalogs;  \\
	4) Selection and projection effects are implemented to achieve the final stage of simulated catalogs ready to compare with observation;\\
	5) Through the lens of Persistent Homology, we also should extract the  topological information from both synthetic and realistic catalogs.\\ 
	Finally, we can make cosmological inferences. Also, in order to create 2-dimensional and pseudo-2-dimensional maps using a 3-dimensional density field, we can either employ the projecting pipeline or decrease the depth size of the simulated box. In both cases, the content of information carried by the Betti numbers can be altered by decreasing the embedding dimension of the input maps. Therefore, our results derived from a 3-dimensional map can be considered as the theoretical estimations of associated quantities. As illustrations, in  \cite{2021A&A...648A..74H, 2022A&A...667A.125H}, the topological measures have been computed for cosmic shear maps and in \cite{2019JCAP...06..019M}, the Minkowski functionals has been used to put constraint on the neutrino mass with weak lensing map.

	To perform a more comprehensive analysis, we suggest the following: In this paper, we used a snapshot at redshift $z=0$, but to achieve more stringent constraints and address the redshift dependence of PH vectorization, a broader range of redshifts should be considered. Additionally, examining redshift dependency offers several cosmological benefits: it allows us to track the evolution of PH vectorization, enabling the inference and discrimination of the impact of different cosmological models based on topological measures throughout the evolution of the universe. In cases where the redshift dependency is not apparent, one can infer the universal behavior of the underlying measures. Consequently, we can extract information from earlier epochs by analyzing the LSS at a later time, ensuring that non-linearities due to time evolution do not obscure primordial effects. Considering additional values of smoothing scales and $N_{bins}$ would be useful for verifying the robustness of PH vectorization relative to the current results. Various coarse-graining pipelines may also be valuable for exploring the multi-scale nature of density fields in the presence of outliers \citep{2024arXiv240313985Y}.  Our approach has potential to offer a fresh perspective on understanding the mass hierarchy of the active neutrinos. Quantifying the information, content usually is carried out by the Fisher forecasts, while an alternative approach so-called simulation-based inference (SBI) known as a likelihood-free analysis is also useful to implement \citep{tejero2020sbi,2016arXiv160506376P,2019MNRAS.488.4440A,2020PNAS..11730055C}. Combining PH measures with other observable quantities can help constrain the uncertainty of cosmological parameters,  which is an important topic for high-precision evaluations.

     \section*{Acknowledgements}

    SMSM and SA appreciate the hospitality of the HECAP section of ICTP where part of this research was completed. We also thank the Quijote team for sharing its simulated data sets and providing extensive instruction on how to utilize the data. SA has received funding/support from the European Union's Horizon 2020 research and innovation program under the Marie Skodowska-Curie Grant Agreement No. 860881- HIDDeN as well as under the Marie Skodowska-Curie Staff Exchange Grant Agreement No. 101086085-ASYMMETRY. The PH calculations were performed by using \texttt{Cubical ripser} \cite{2020arXiv200512692K}, and  \texttt{GUDHI} \cite{maria2014gudhi}  python packages. Also, the Fisher forecast plots are provided by utilizing \texttt{Getdist} python package \cite{2019arXiv191013970L}. Finally, we thank Syeda Aliya Batool for reading the manuscript and giving her constructive comments.

    \section*{Data Availability}

    The new data generated and the computational program underlying this article will be shared on reasonable request to the corresponding author.



\begin{thebibliography}{}
\makeatletter
\relax
\def\mn@urlcharsother{\let\do\@makeother \do\$\do\&\do\#\do\^\do\_\do\%\do\~}
\def\mn@doi{\begingroup\mn@urlcharsother \@ifnextchar [ {\mn@doi@}
  {\mn@doi@[]}}
\def\mn@doi@[#1]#2{\def\@tempa{#1}\ifx\@tempa\@empty \href
  {http://dx.doi.org/#2} {doi:#2}\else \href {http://dx.doi.org/#2} {#1}\fi
  \endgroup}
\def\mn@eprint#1#2{\mn@eprint@#1:#2::\@nil}
\def\mn@eprint@arXiv#1{\href {http://arxiv.org/abs/#1} {{\tt arXiv:#1}}}
\def\mn@eprint@dblp#1{\href {http://dblp.uni-trier.de/rec/bibtex/#1.xml}
  {dblp:#1}}
\def\mn@eprint@#1:#2:#3:#4\@nil{\def\@tempa {#1}\def\@tempb {#2}\def\@tempc
  {#3}\ifx \@tempc \@empty \let \@tempc \@tempb \let \@tempb \@tempa \fi \ifx
  \@tempb \@empty \def\@tempb {arXiv}\fi \@ifundefined
  {mn@eprint@\@tempb}{\@tempb:\@tempc}{\expandafter \expandafter \csname
  mn@eprint@\@tempb\endcsname \expandafter{\@tempc}}}

\bibitem[\protect\citeauthoryear{Abazajian}{Abazajian}{2017}]{Abazajian:2017tcc}
Abazajian K.~N.,  2017, \mn@doi [Phys. Rept.] {10.1016/j.physrep.2017.10.003},
  711-712, 1

\bibitem[\protect\citeauthoryear{Abazajian et~al.,}{Abazajian
  et~al.}{2016}]{CMB-S4:2016ple}
Abazajian K.~N.,  et~al., 2016, arXiv preprint arXiv:1610.02743

\bibitem[\protect\citeauthoryear{{Abazajian} et~al.,}{{Abazajian}
  et~al.}{2019}]{2019arXiv190704473A}
{Abazajian} K.,  et~al., 2019, \mn@doi [arXiv e-prints]
  {10.48550/arXiv.1907.04473}, \href
  {https://ui.adsabs.harvard.edu/abs/2019arXiv190704473A} {p. arXiv:1907.04473}

\bibitem[\protect\citeauthoryear{Abbott et~al.,}{Abbott
  et~al.}{2023}]{Kilo-DegreeSurvey:2023gfr}
Abbott T.,  et~al., 2023, The Open Journal of Astrophysics, 6, 1

\bibitem[\protect\citeauthoryear{Abell et~al.,}{Abell
  et~al.}{2009}]{LSSTScience:2009jmu}
Abell P.~A.,  et~al., 2009, arXiv preprint arXiv:0912.0201

\bibitem[\protect\citeauthoryear{{Abitbol} et~al.,}{{Abitbol}
  et~al.}{2017}]{2017arXiv170602464A}
{Abitbol} M.~H.,  et~al., 2017, \mn@doi [arXiv e-prints]
  {10.48550/arXiv.1706.02464}, \href
  {https://ui.adsabs.harvard.edu/abs/2017arXiv170602464A} {p. arXiv:1706.02464}

\bibitem[\protect\citeauthoryear{Adams et~al.,}{Adams
  et~al.}{2017}]{adams2017persistence}
Adams H.,  et~al., 2017, Journal of Machine Learning Research, 18, 1

\bibitem[\protect\citeauthoryear{Ade et~al.}{Ade
  et~al.}{2019}]{SimonsObservatory:2018koc}
Ade P.,  et~al., 2019, \mn@doi [JCAP] {10.1088/1475-7516/2019/02/056}, 02, 056

\bibitem[\protect\citeauthoryear{{Agarwal} \& {Feldman}}{{Agarwal} \&
  {Feldman}}{2011}]{2011MNRAS.410.1647A}
{Agarwal} S.,  {Feldman} H.~A.,  2011, \mn@doi [\mnras]
  {10.1111/j.1365-2966.2010.17546.x}, \href
  {https://ui.adsabs.harvard.edu/abs/2011MNRAS.410.1647A} {410, 1647}

\bibitem[\protect\citeauthoryear{Aghanim et~al.,}{Aghanim
  et~al.}{2020}]{aghanim2020planckvi}
Aghanim N.,  et~al., 2020, Astronomy \& Astrophysics, 641, A6

\bibitem[\protect\citeauthoryear{Alam et~al.}{Alam
  et~al.}{2021}]{eBOSS:2020yzd}
Alam S.,  et~al., 2021, \mn@doi [Phys. Rev. D] {10.1103/PhysRevD.103.083533},
  103, 083533

\bibitem[\protect\citeauthoryear{Alesker}{Alesker}{1999}]{Alesker1999}
Alesker S.,  1999, Geometriae Dedicata, 74, 241

\bibitem[\protect\citeauthoryear{{Alsing}, {Charnock}, {Feeney}  \&
  {Wandelt}}{{Alsing} et~al.}{2019}]{2019MNRAS.488.4440A}
{Alsing} J.,  {Charnock} T.,  {Feeney} S.,   {Wandelt} B.,  2019, \mn@doi
  [Monthly Notices of the RAS] {10.1093/mnras/stz1960}, \href
  {https://ui.adsabs.harvard.edu/abs/2019MNRAS.488.4440A} {488, 4440}

\bibitem[\protect\citeauthoryear{{Banerjee} \& {Dalal}}{{Banerjee} \&
  {Dalal}}{2016}]{2016JCAP...11..015B}
{Banerjee} A.,  {Dalal} N.,  2016, \mn@doi [\jcap]
  {10.1088/1475-7516/2016/11/015}, \href
  {https://ui.adsabs.harvard.edu/abs/2016JCAP...11..015B} {2016, 015}

\bibitem[\protect\citeauthoryear{{Bayer} et~al.,}{{Bayer}
  et~al.}{2021}]{2021ApJ...919...24B}
{Bayer} A.~E.,  et~al., 2021, \mn@doi [Astrophysical Journal]
  {10.3847/1538-4357/ac0e91}, \href
  {https://ui.adsabs.harvard.edu/abs/2021ApJ...919...24B} {919, 24}

\bibitem[\protect\citeauthoryear{{Bayer}, {Banerjee}  \& {Seljak}}{{Bayer}
  et~al.}{2022}]{2022PhRvD.105l3510B}
{Bayer} A.~E.,  {Banerjee} A.,   {Seljak} U.,  2022, \mn@doi [\prd]
  {10.1103/PhysRevD.105.123510}, \href
  {https://ui.adsabs.harvard.edu/abs/2022PhRvD.105l3510B} {105, 123510}

\bibitem[\protect\citeauthoryear{Beisbart, Dahlke, Mecke  \& Wagner}{Beisbart
  et~al.}{2002}]{Beisbart2002}
Beisbart C.,  Dahlke R.,  Mecke K.,   Wagner H.,  2002, Morphology of Condensed
  Matter: Physics and Geometry of Spatially Complex Systems, pp 238--260

\bibitem[\protect\citeauthoryear{{Bernardeau}}{{Bernardeau}}{1994}]{1994A&A...291..697B}
{Bernardeau} F.,  1994, \mn@doi [Astronomy and Astrophysics]
  {10.48550/arXiv.astro-ph/9403020}, \href
  {https://ui.adsabs.harvard.edu/abs/1994A&A...291..697B} {291, 697}

\bibitem[\protect\citeauthoryear{Biagetti, Cole  \& Shiu}{Biagetti
  et~al.}{2021}]{biagetti2021persistence}
Biagetti M.,  Cole A.,   Shiu G.,  2021, Journal of Cosmology and Astroparticle
  Physics, 2021, 061

\bibitem[\protect\citeauthoryear{{Biagetti}, {Calles}, {Castiblanco}, {Cole}
  \& {Nore{\~n}a}}{{Biagetti} et~al.}{2022}]{2022JCAP...10..002B}
{Biagetti} M.,  {Calles} J.,  {Castiblanco} L.,  {Cole} A.,   {Nore{\~n}a} J.,
  2022, \mn@doi [Journal of Cosmology and Astroparticle Physics]
  {10.1088/1475-7516/2022/10/002}, \href
  {https://ui.adsabs.harvard.edu/abs/2022JCAP...10..002B} {2022, 002}

\bibitem[\protect\citeauthoryear{Brout et~al.}{Brout
  et~al.}{2022}]{Brout:2022vxf}
Brout D.,  et~al., 2022, \mn@doi [Astrophys. J.] {10.3847/1538-4357/ac8e04},
  938, 110

\bibitem[\protect\citeauthoryear{Capozzi, Di~Valentino, Lisi, Marrone,
  Melchiorri  \& Palazzo}{Capozzi et~al.}{2017}]{Capozzi:2017ipn}
Capozzi F.,  Di~Valentino E.,  Lisi E.,  Marrone A.,  Melchiorri A.,   Palazzo
  A.,  2017, \mn@doi [Phys. Rev. D] {10.1103/PhysRevD.95.096014}, 95, 096014

\bibitem[\protect\citeauthoryear{Capozzi, Lisi, Marrone  \& Palazzo}{Capozzi
  et~al.}{2018}]{Capozzi:2018ubv}
Capozzi F.,  Lisi E.,  Marrone A.,   Palazzo A.,  2018, \mn@doi [Prog. Part.
  Nucl. Phys.] {10.1016/j.ppnp.2018.05.005}, 102, 48

\bibitem[\protect\citeauthoryear{{Castorina}, {Sefusatti}, {Sheth},
  {Villaescusa-Navarro}  \& {Viel}}{{Castorina}
  et~al.}{2014}]{2014JCAP...02..049C}
{Castorina} E.,  {Sefusatti} E.,  {Sheth} R.~K.,  {Villaescusa-Navarro} F.,
  {Viel} M.,  2014, \mn@doi [Journal of Cosmology and Astroparticle Physics]
  {10.1088/1475-7516/2014/02/049}, \href
  {https://ui.adsabs.harvard.edu/abs/2014JCAP...02..049C} {2014, 049}

\bibitem[\protect\citeauthoryear{Chiang, Hu, Li  \& Loverde}{Chiang
  et~al.}{2017}]{Chiang2017ScaledependentBA}
Chiang C.-T.,  Hu W.,  Li Y.,   Loverde M.,  2017, Physical Review D

\bibitem[\protect\citeauthoryear{{Chiang}, {LoVerde}  \&
  {Villaescusa-Navarro}}{{Chiang} et~al.}{2019}]{Chiang2018FirstDO}
{Chiang} C.-T.,  {LoVerde} M.,   {Villaescusa-Navarro} F.,  2019, \mn@doi
  [Physical review letters] {10.1103/PhysRevLett.122.041302}, \href
  {https://ui.adsabs.harvard.edu/abs/2019PhRvL.122d1302C} {122, 041302}

\bibitem[\protect\citeauthoryear{{Chudaykin} \& {Ivanov}}{{Chudaykin} \&
  {Ivanov}}{2019}]{2019JCAP...11..034C}
{Chudaykin} A.,  {Ivanov} M.~M.,  2019, \mn@doi [Journal of Cosmology and
  Astroparticle Physics] {10.1088/1475-7516/2019/11/034}, \href
  {https://ui.adsabs.harvard.edu/abs/2019JCAP...11..034C} {2019, 034}

\bibitem[\protect\citeauthoryear{{Cisewski-Kehe}, {Fasy}, {Hellwing}, {Lovell},
  {Drozda}  \& {Wu}}{{Cisewski-Kehe} et~al.}{2022}]{Cisewski-Kehe2022PhRvD}
{Cisewski-Kehe} J.,  {Fasy} B.~T.,  {Hellwing} W.,  {Lovell} M.~R.,  {Drozda}
  P.,   {Wu} M.,  2022, \mn@doi [Physical Review D]
  {10.1103/PhysRevD.106.023521}, \href
  {https://ui.adsabs.harvard.edu/abs/2022PhRvD.106b3521C} {106, 023521}

\bibitem[\protect\citeauthoryear{{Cole} \& {Shiu}}{{Cole} \&
  {Shiu}}{2018}]{2018JCAP...03..025C}
{Cole} A.,  {Shiu} G.,  2018, \mn@doi [Journal of Cosmology and Astroparticle
  Physics] {10.1088/1475-7516/2018/03/025}, \href
  {https://ui.adsabs.harvard.edu/abs/2018JCAP...03..025C} {2018, 025}

\bibitem[\protect\citeauthoryear{Contarini, Marulli, Moscardini, Veropalumbo,
  Giocoli  \& Baldi}{Contarini et~al.}{2021}]{Contarini2020CosmicVI}
Contarini S.,  Marulli F.,  Moscardini L.,  Veropalumbo A.,  Giocoli C.,
  Baldi M.,  2021, \mn@doi [MNRAS] {10.1093/mnras/stab1112}, 504, 5021

\bibitem[\protect\citeauthoryear{{Cranmer}, {Brehmer}  \& {Louppe}}{{Cranmer}
  et~al.}{2020}]{2020PNAS..11730055C}
{Cranmer} K.,  {Brehmer} J.,   {Louppe} G.,  2020, \mn@doi [Proceedings of the
  National Academy of Science] {10.1073/pnas.1912789117}, \href
  {https://ui.adsabs.harvard.edu/abs/2020PNAS..11730055C} {117, 30055}

\bibitem[\protect\citeauthoryear{{DESI Collaboration} et~al.,}{{DESI
  Collaboration} et~al.}{2024}]{2024arXiv240403002D}
{DESI Collaboration} et~al., 2024, \mn@doi [arXiv e-prints]
  {10.48550/arXiv.2404.03002}, \href
  {https://ui.adsabs.harvard.edu/abs/2024arXiv240403002D} {p. arXiv:2404.03002}

\bibitem[\protect\citeauthoryear{{Dark Energy Survey and Kilo-Degree Survey
  Collaboration} et~al.,}{{Dark Energy Survey and Kilo-Degree Survey
  Collaboration} et~al.}{2023}]{46ddc1eeccc74c52ba263dcebe509c3a}
{Dark Energy Survey and Kilo-Degree Survey Collaboration} et~al., 2023, \mn@doi
  [The Open Journal of Astrophysics] {10.21105/astro.2305.17173}, 6, 1

\bibitem[\protect\citeauthoryear{{Desjacques}, {Jeong}  \&
  {Schmidt}}{{Desjacques} et~al.}{2018}]{2018PhR...733....1D}
{Desjacques} V.,  {Jeong} D.,   {Schmidt} F.,  2018, \mn@doi [Physics Reports]
  {10.1016/j.physrep.2017.12.002}, \href
  {https://ui.adsabs.harvard.edu/abs/2018PhR...733....1D} {733, 1}

\bibitem[\protect\citeauthoryear{Dey \& Wang}{Dey \&
  Wang}{2022}]{dey2022computational}
Dey T.~K.,  Wang Y.,  2022, Computational topology for data analysis.
Cambridge University Press

\bibitem[\protect\citeauthoryear{{Dodelson} \& {Vesterinen}}{{Dodelson} \&
  {Vesterinen}}{2009}]{2009PhRvL.103q1301D}
{Dodelson} S.,  {Vesterinen} M.,  2009, \mn@doi [Physical Review Letters]
  {10.1103/PhysRevLett.103.171301}, \href
  {https://ui.adsabs.harvard.edu/abs/2009PhRvL.103q1301D} {103, 171301}

\bibitem[\protect\citeauthoryear{Dodelson, Gates  \& Stebbins}{Dodelson
  et~al.}{1996}]{Dodelson:1995es}
Dodelson S.,  Gates E.,   Stebbins A.,  1996, \mn@doi [Astrophys. J.]
  {10.1086/177581}, 467, 10

\bibitem[\protect\citeauthoryear{Edelsbrunner \& Harer}{Edelsbrunner \&
  Harer}{2022}]{edelsbrunner2022computational}
Edelsbrunner H.,  Harer J.~L.,  2022, Computational topology: an introduction.
American Mathematical Society

\bibitem[\protect\citeauthoryear{Edelsbrunner \& M{\"u}cke}{Edelsbrunner \&
  M{\"u}cke}{1994}]{edelsbrunner1994three}
Edelsbrunner H.,  M{\"u}cke E.~P.,  1994, ACM Transactions On Graphics (TOG),
  13, 43

\bibitem[\protect\citeauthoryear{Edelsbrunner, Letscher  \&
  Zomorodian}{Edelsbrunner et~al.}{2000}]{edelsbrunner2000proceedings}
Edelsbrunner H.,  Letscher D.,   Zomorodian A.,  2000, Proceedings 41st annual
  symposium on foundations of computer science

\bibitem[\protect\citeauthoryear{{Elbers} \& {van de Weygaert}}{{Elbers} \&
  {van de Weygaert}}{2023}]{2023MNRAS.520.2709E}
{Elbers} W.,  {van de Weygaert} R.,  2023, \mn@doi [Monthly Notices of the
  Royal Astronomical Society] {10.1093/mnras/stad120}, \href
  {https://ui.adsabs.harvard.edu/abs/2023MNRAS.520.2709E} {520, 2709}

\bibitem[\protect\citeauthoryear{Esteban, Gonzalez-Garcia, Maltoni, Schwetz  \&
  Zhou}{Esteban et~al.}{2020}]{Esteban:2020cvm}
Esteban I.,  Gonzalez-Garcia M.~C.,  Maltoni M.,  Schwetz T.,   Zhou A.,  2020,
  \mn@doi [JHEP] {10.1007/JHEP09(2020)178}, 09, 178

\bibitem[\protect\citeauthoryear{{Euclid Collaboration} et~al.,}{{Euclid
  Collaboration} et~al.}{2024}]{Euclid:2024imf}
{Euclid Collaboration} et~al., 2024, \mn@doi [arXiv e-prints]
  {10.48550/arXiv.2405.06047}, \href
  {https://ui.adsabs.harvard.edu/abs/2024arXiv240506047E} {p. arXiv:2405.06047}

\bibitem[\protect\citeauthoryear{{Feldbrugge}, {van Engelen}, {van de
  Weygaert}, {Pranav}  \& {Vegter}}{{Feldbrugge}
  et~al.}{2019}]{feldbrugge2019stochastic}
{Feldbrugge} J.,  {van Engelen} M.,  {van de Weygaert} R.,  {Pranav} P.,
  {Vegter} G.,  2019, \mn@doi [Journal of Cosmology and Astroparticle Physics]
  {10.1088/1475-7516/2019/09/052}, \href
  {https://ui.adsabs.harvard.edu/abs/2019JCAP...09..052F} {2019, 052}

\bibitem[\protect\citeauthoryear{{Gong} et~al.,}{{Gong}
  et~al.}{2019}]{2019ApJ883}
{Gong} Y.,  et~al., 2019, \mn@doi [Astrophysical Journal]
  {10.3847/1538-4357/ab391e}, \href
  {https://ui.adsabs.harvard.edu/abs/2019ApJ...883..203G} {883, 203}

\bibitem[\protect\citeauthoryear{{Gott}, {Melott}  \& {Dickinson}}{{Gott}
  et~al.}{1986}]{1986ApJ...306..341G}
{Gott} J.~Richard I.,  {Melott} A.~L.,   {Dickinson} M.,  1986, \mn@doi [\apj]
  {10.1086/164347}, \href
  {https://ui.adsabs.harvard.edu/abs/1986ApJ...306..341G} {306, 341}

\bibitem[\protect\citeauthoryear{{Gott} J.~Richard et~al.,}{{Gott}
  et~al.}{1989}]{1989ApJ...340..625G}
{Gott} J.~Richard I.,  et~al., 1989, \mn@doi [\apj] {10.1086/167425}, \href
  {https://ui.adsabs.harvard.edu/abs/1989ApJ...340..625G} {340, 625}

\bibitem[\protect\citeauthoryear{{Hahn}, {Villaescusa-Navarro}, {Castorina}  \&
  {Scoccimarro}}{{Hahn} et~al.}{2020}]{2020JCAP...03..040H}
{Hahn} C.,  {Villaescusa-Navarro} F.,  {Castorina} E.,   {Scoccimarro} R.,
  2020, \mn@doi [Journal of Cosmology and Astroparticle Physics]
  {10.1088/1475-7516/2020/03/040}, \href
  {https://ui.adsabs.harvard.edu/abs/2020JCAP...03..040H} {2020, 040}

\bibitem[\protect\citeauthoryear{{Hamilton}, {Gott}  \& {Weinberg}}{{Hamilton}
  et~al.}{1986}]{1986ApJ...309....1H}
{Hamilton} A.~J.~S.,  {Gott} J.~Richard I.,   {Weinberg} D.,  1986, \mn@doi
  [\apj] {10.1086/164571}, \href
  {https://ui.adsabs.harvard.edu/abs/1986ApJ...309....1H} {309, 1}

\bibitem[\protect\citeauthoryear{{Hartlap}, {Simon}  \& {Schneider}}{{Hartlap}
  et~al.}{2007}]{2007A&A...464..399H}
{Hartlap} J.,  {Simon} P.,   {Schneider} P.,  2007, \mn@doi [\aap]
  {10.1051/0004-6361:20066170}, \href
  {https://ui.adsabs.harvard.edu/abs/2007A&A...464..399H} {464, 399}

\bibitem[\protect\citeauthoryear{{Heydenreich}, {Br{\"u}ck}  \&
  {Harnois-D{\'e}raps}}{{Heydenreich} et~al.}{2021}]{2021A&A...648A..74H}
{Heydenreich} S.,  {Br{\"u}ck} B.,   {Harnois-D{\'e}raps} J.,  2021, \mn@doi
  [Astronomy and Astrophysics] {10.1051/0004-6361/202039048}, \href
  {https://ui.adsabs.harvard.edu/abs/2021A&A...648A..74H} {648, A74}

\bibitem[\protect\citeauthoryear{{Heydenreich}, {Br{\"u}ck}, {Burger},
  {Harnois-D{\'e}raps}, {Unruh}, {Castro}, {Dolag}  \&
  {Martinet}}{{Heydenreich} et~al.}{2022}]{2022A&A...667A.125H}
{Heydenreich} S.,  {Br{\"u}ck} B.,  {Burger} P.,  {Harnois-D{\'e}raps} J.,
  {Unruh} S.,  {Castro} T.,  {Dolag} K.,   {Martinet} N.,  2022, \mn@doi
  [Astronomy and Astrophysics] {10.1051/0004-6361/202243868}, \href
  {https://ui.adsabs.harvard.edu/abs/2022A&A...667A.125H} {667, A125}

\bibitem[\protect\citeauthoryear{{Hoekstra} \& {Jain}}{{Hoekstra} \&
  {Jain}}{2008}]{2008ARNPS..58...99H}
{Hoekstra} H.,  {Jain} B.,  2008, \mn@doi [Annual Review of Nuclear and
  Particle Science] {10.1146/annurev.nucl.58.110707.171151}, \href
  {https://ui.adsabs.harvard.edu/abs/2008ARNPS..58...99H} {58, 99}

\bibitem[\protect\citeauthoryear{Hug, Schneider  \& Schuster}{Hug
  et~al.}{2007}]{Hug2007TheSO}
Hug D.,  Schneider R.,   Schuster R.,  2007, St Petersburg Mathematical
  Journal, 19, 137

\bibitem[\protect\citeauthoryear{{Ichiki} \& {Takada}}{{Ichiki} \&
  {Takada}}{2012}]{2012PhRvD..85f3521I}
{Ichiki} K.,  {Takada} M.,  2012, \mn@doi [Physical Review D]
  {10.1103/PhysRevD.85.063521}, \href
  {https://ui.adsabs.harvard.edu/abs/2012PhRvD..85f3521I} {85, 063521}

\bibitem[\protect\citeauthoryear{{Jenkins}, {Frenk}, {White}, {Colberg},
  {Cole}, {Evrard}, {Couchman}  \& {Yoshida}}{{Jenkins}
  et~al.}{2001}]{2001MNRAS.321..372J}
{Jenkins} A.,  {Frenk} C.~S.,  {White} S.~D.~M.,  {Colberg} J.~M.,  {Cole} S.,
  {Evrard} A.~E.,  {Couchman} H.~M.~P.,   {Yoshida} N.,  2001, \mn@doi [Monthly
  Notices of the Royal Astronomical Society]
  {10.1046/j.1365-8711.2001.04029.x}, \href
  {https://ui.adsabs.harvard.edu/abs/2001MNRAS.321..372J} {321, 372}

\bibitem[\protect\citeauthoryear{{Kaji}, {Sudo}  \& {Ahara}}{{Kaji}
  et~al.}{2020}]{2020arXiv200512692K}
{Kaji} S.,  {Sudo} T.,   {Ahara} K.,  2020, \mn@doi [arXiv e-prints]
  {10.48550/arXiv.2005.12692}, \href
  {https://ui.adsabs.harvard.edu/abs/2020arXiv200512692K} {p. arXiv:2005.12692}

\bibitem[\protect\citeauthoryear{{Kanafi} \& {Movahed}}{{Kanafi} \&
  {Movahed}}{2024}]{2024ApJ...963...31K}
{Kanafi} M.~H.~J.,  {Movahed} S.~M.~S.,  2024, \mn@doi [\apj]
  {10.3847/1538-4357/ad1880}, \href
  {https://ui.adsabs.harvard.edu/abs/2024ApJ...963...31K} {963, 31}

\bibitem[\protect\citeauthoryear{Kelly, Machado, Parke, Perez-Gonzalez  \&
  Funchal}{Kelly et~al.}{2021}]{Kelly:2020fkv}
Kelly K.~J.,  Machado P. A.~N.,  Parke S.~J.,  Perez-Gonzalez Y.~F.,   Funchal
  R.~Z.,  2021, \mn@doi [Phys. Rev. D] {10.1103/PhysRevD.103.013004}, 103,
  013004

\bibitem[\protect\citeauthoryear{{Kreisch}, {Pisani}, {Carbone}, {Liu},
  {Hawken}, {Massara}, {Spergel}  \& {Wandelt}}{{Kreisch}
  et~al.}{2019}]{2019MNRAS.488.4413K}
{Kreisch} C.~D.,  {Pisani} A.,  {Carbone} C.,  {Liu} J.,  {Hawken} A.~J.,
  {Massara} E.,  {Spergel} D.~N.,   {Wandelt} B.~D.,  2019, \mn@doi [Monthly
  Notices of the RAS] {10.1093/mnras/stz1944}, \href
  {https://ui.adsabs.harvard.edu/abs/2019MNRAS.488.4413K} {488, 4413}

\bibitem[\protect\citeauthoryear{Laureijs et~al.,}{Laureijs
  et~al.}{2011}]{EUCLID:2011zbd}
Laureijs R.,  et~al., 2011, arXiv preprint arXiv:1110.3193

\bibitem[\protect\citeauthoryear{{Lazeyras}, {Villaescusa-Navarro}  \&
  {Viel}}{{Lazeyras} et~al.}{2021}]{2021JCAP...03..022L}
{Lazeyras} T.,  {Villaescusa-Navarro} F.,   {Viel} M.,  2021, \mn@doi [Journal
  of Cosmology and Astroparticle Physics] {10.1088/1475-7516/2021/03/022},
  \href {https://ui.adsabs.harvard.edu/abs/2021JCAP...03..022L} {2021, 022}

\bibitem[\protect\citeauthoryear{{Lesgourgues} \& {Pastor}}{{Lesgourgues} \&
  {Pastor}}{2006}]{Lesgourgues:2006nd}
{Lesgourgues} J.,  {Pastor} S.,  2006, \mn@doi [Physics Reports]
  {10.1016/j.physrep.2006.04.001}, \href
  {https://ui.adsabs.harvard.edu/abs/2006PhR...429..307L} {429, 307}

\bibitem[\protect\citeauthoryear{Lesgourgues \& Pastor}{Lesgourgues \&
  Pastor}{2012}]{2012arXiv1212.6154L}
Lesgourgues J.,  Pastor S.,  2012, \mn@doi [Adv. High Energy Phys.]
  {10.1155/2012/608515}, 2012, 608515

\bibitem[\protect\citeauthoryear{Lesgourgues, Mangano, Miele  \&
  Pastor}{Lesgourgues et~al.}{2013}]{lesgourgues2013neutrino}
Lesgourgues J.,  Mangano G.,  Miele G.,   Pastor S.,  2013, Neutrino cosmology.
Cambridge University Press

\bibitem[\protect\citeauthoryear{{Lewis}}{{Lewis}}{2019}]{2019arXiv191013970L}
{Lewis} A.,  2019, \mn@doi [arXiv e-prints] {10.48550/arXiv.1910.13970}, \href
  {https://ui.adsabs.harvard.edu/abs/2019arXiv191013970L} {p. arXiv:1910.13970}

\bibitem[\protect\citeauthoryear{{Liu}, {Yu}, {Yu}  \& {Zhang}}{{Liu}
  et~al.}{2020}]{liu2020neutrino}
{Liu} Y.,  {Yu} Y.,  {Yu} H.-R.,   {Zhang} P.,  2020, \mn@doi [Physical Review
  D] {10.1103/PhysRevD.101.063515}, \href
  {https://ui.adsabs.harvard.edu/abs/2020PhRvD.101f3515L} {101, 063515}

\bibitem[\protect\citeauthoryear{{Liu}, {Jiang}  \& {Fang}}{{Liu}
  et~al.}{2022}]{Liu2022ProbingMN}
{Liu} W.,  {Jiang} A.,   {Fang} W.,  2022, \mn@doi [Journal of Cosmology and
  Astroparticle Physics] {10.1088/1475-7516/2022/07/045}, \href
  {https://ui.adsabs.harvard.edu/abs/2022JCAP...07..045L} {2022, 045}

\bibitem[\protect\citeauthoryear{Liu, Jiang  \& Fang}{Liu
  et~al.}{2023}]{Liu2023ProbingMN}
Liu W.,  Jiang A.,   Fang W.,  2023, \mn@doi [Journal of Cosmology and
  Astroparticle Physics] {10.1088/1475-7516/2023/09/037}, \href
  {https://ui.adsabs.harvard.edu/abs/2023JCAP...09..037L} {2023, 037}

\bibitem[\protect\citeauthoryear{{Lucie-Smith}, {Barreira}  \&
  {Schmidt}}{{Lucie-Smith} et~al.}{2023}]{2023MNRAS.524.1746L}
{Lucie-Smith} L.,  {Barreira} A.,   {Schmidt} F.,  2023, \mn@doi [Monthly
  Notices of the Royal Astronomical Society] {10.1093/mnras/stad2003}, \href
  {https://ui.adsabs.harvard.edu/abs/2023MNRAS.524.1746L} {524, 1746}

\bibitem[\protect\citeauthoryear{Madhavacheril et~al.}{Madhavacheril
  et~al.}{2024}]{ACT:2023kun}
Madhavacheril M.~S.,  et~al., 2024, \mn@doi [Astrophys. J.]
  {10.3847/1538-4357/acff5f}, 962, 113

\bibitem[\protect\citeauthoryear{Maria, Boissonnat, Glisse  \& Yvinec}{Maria
  et~al.}{2014}]{maria2014gudhi}
Maria C.,  Boissonnat J.-D.,  Glisse M.,   Yvinec M.,  2014, in Mathematical
  Software--ICMS 2014: 4th International Congress, Seoul, South Korea, August
  5-9, 2014. Proceedings 4. pp 167--174

\bibitem[\protect\citeauthoryear{{Marques}, {Liu}, {Zorrilla Matilla},
  {Haiman}, {Bernui}  \& {Novaes}}{{Marques}
  et~al.}{2019}]{2019JCAP...06..019M}
{Marques} G.~A.,  {Liu} J.,  {Zorrilla Matilla} J.~M.,  {Haiman} Z.,  {Bernui}
  A.,   {Novaes} C.~P.,  2019, \mn@doi [Journal of Cosmology and Astroparticle
  Physics] {10.1088/1475-7516/2019/06/019}, \href
  {https://ui.adsabs.harvard.edu/abs/2019JCAP...06..019M} {2019, 019}

\bibitem[\protect\citeauthoryear{{Marulli}, {Carbone}, {Viel}, {Moscardini}  \&
  {Cimatti}}{{Marulli} et~al.}{2011}]{2011MNRAS.418..346M}
{Marulli} F.,  {Carbone} C.,  {Viel} M.,  {Moscardini} L.,   {Cimatti} A.,
  2011, \mn@doi [Monthly Notices of the RAS]
  {10.1111/j.1365-2966.2011.19488.x}, \href
  {https://ui.adsabs.harvard.edu/abs/2011MNRAS.418..346M} {418, 346}

\bibitem[\protect\citeauthoryear{{Masoomy}, {Askari}, {Najafi}  \&
  {Movahed}}{{Masoomy} et~al.}{2021}]{masoomy2021persistent}
{Masoomy} H.,  {Askari} B.,  {Najafi} M.~N.,   {Movahed} S.~M.~S.,  2021,
  \mn@doi [Physical Review E] {10.1103/PhysRevE.104.034116}, \href
  {https://ui.adsabs.harvard.edu/abs/2021PhRvE.104c4116M} {104, 034116}

\bibitem[\protect\citeauthoryear{{Masoomy}, {Tajik}  \& {Movahed}}{{Masoomy}
  et~al.}{2022}]{2022PhRvE.106f4115M}
{Masoomy} H.,  {Tajik} S.,   {Movahed} S.~M.~S.,  2022, \mn@doi [Physical
  Review E] {10.1103/PhysRevE.106.064115}, \href
  {https://ui.adsabs.harvard.edu/abs/2022PhRvE.106f4115M} {106, 064115}

\bibitem[\protect\citeauthoryear{{Massara}, {Villaescusa-Navarro}, {Viel}  \&
  {Sutter}}{{Massara} et~al.}{2015}]{2015JCAP...11..018M}
{Massara} E.,  {Villaescusa-Navarro} F.,  {Viel} M.,   {Sutter} P.~M.,  2015,
  \mn@doi [\jcap] {10.1088/1475-7516/2015/11/018}, \href
  {https://ui.adsabs.harvard.edu/abs/2015JCAP...11..018M} {2015, 018}

\bibitem[\protect\citeauthoryear{{Massara}, {Villaescusa-Navarro}, {Ho},
  {Dalal}  \& {Spergel}}{{Massara} et~al.}{2021}]{2021PhRvL.126a1301M}
{Massara} E.,  {Villaescusa-Navarro} F.,  {Ho} S.,  {Dalal} N.,   {Spergel}
  D.~N.,  2021, \mn@doi [Physical Review Letters]
  {10.1103/PhysRevLett.126.011301}, \href
  {https://ui.adsabs.harvard.edu/abs/2021PhRvL.126a1301M} {126, 011301}

\bibitem[\protect\citeauthoryear{{Matsubara}}{{Matsubara}}{2003}]{matsubara2003statistics}
{Matsubara} T.,  2003, \mn@doi [Astrophysical Journal] {10.1086/345521}, \href
  {https://ui.adsabs.harvard.edu/abs/2003ApJ...584....1M} {584, 1}

\bibitem[\protect\citeauthoryear{{Matsubara}, {Hikage}  \&
  {Kuriki}}{{Matsubara} et~al.}{2022}]{matsubara2022minkowski}
{Matsubara} T.,  {Hikage} C.,   {Kuriki} S.,  2022, \mn@doi [Physical Review D]
  {10.1103/PhysRevD.105.023527}, \href
  {https://ui.adsabs.harvard.edu/abs/2022PhRvD.105b3527M} {105, 023527}

\bibitem[\protect\citeauthoryear{McMullen}{McMullen}{1997}]{McMullen1997}
McMullen P.,  1997. \url {https://api.semanticscholar.org/CorpusID:117290019}

\bibitem[\protect\citeauthoryear{{Moon}, {Rossi}  \& {Yu}}{{Moon}
  et~al.}{2023}]{Moon2023ApJS}
{Moon} J.,  {Rossi} G.,   {Yu} H.,  2023, \mn@doi [Astrophysical Journal,
  Supplement] {10.3847/1538-4365/aca32a}, \href
  {https://ui.adsabs.harvard.edu/abs/2023ApJS..264...26M} {264, 26}

\bibitem[\protect\citeauthoryear{Munkres}{Munkres}{2018}]{munkres2018elements}
Munkres J.~R.,  2018, Elements of algebraic topology.
CRC Press

\bibitem[\protect\citeauthoryear{{Murata} \& {Matsubara}}{{Murata} \&
  {Matsubara}}{2007}]{2007PASJ...59...73M}
{Murata} Y.,  {Matsubara} T.,  2007, \mn@doi [\pasj] {10.1093/pasj/59.1.73},
  \href {https://ui.adsabs.harvard.edu/abs/2007PASJ...59...73M} {59, 73}

\bibitem[\protect\citeauthoryear{Nakahara}{Nakahara}{2003}]{nakahara2003geometry}
Nakahara M.,  2003, Geometry, topology and physics.
CRC Press

\bibitem[\protect\citeauthoryear{Otter, Porter, Tillmann, Grindrod  \&
  Harrington}{Otter et~al.}{2017}]{otter2017roadmap}
Otter N.,  Porter M.~A.,  Tillmann U.,  Grindrod P.,   Harrington H.~A.,  2017,
  EPJ Data Science, 6, 17

\bibitem[\protect\citeauthoryear{Pan et~al.,}{Pan et~al.}{2023}]{SPT:2023jql}
Pan Z.,  et~al., 2023, \mn@doi [Phys. Rev. D] {10.1103/PhysRevD.108.122005},
  108, 122005

\bibitem[\protect\citeauthoryear{Papamakarios \& Murray}{Papamakarios \&
  Murray}{2016}]{2016arXiv160506376P}
Papamakarios G.,  Murray I.,  2016, Advances in neural information processing
  systems, 29

\bibitem[\protect\citeauthoryear{{Parimbelli} et~al.,}{{Parimbelli}
  et~al.}{2021}]{2021JCAP...01..009P}
{Parimbelli} G.,  et~al., 2021, \mn@doi [Journal of Cosmology and Astroparticle
  Physics] {10.1088/1475-7516/2021/01/009}, \href
  {https://ui.adsabs.harvard.edu/abs/2021JCAP...01..009P} {2021, 009}

\bibitem[\protect\citeauthoryear{Pereira \& de Mello}{Pereira \&
  de~Mello}{2015}]{pereira2015persistent}
Pereira C.~M.,  de Mello R.~F.,  2015, Expert Systems with Applications, 42,
  6026

\bibitem[\protect\citeauthoryear{{Philcox} \& {Torquato}}{{Philcox} \&
  {Torquato}}{2023}]{2023PhRvX..13a1038P}
{Philcox} O. H.~E.,  {Torquato} S.,  2023, \mn@doi [Physical Review X]
  {10.1103/PhysRevX.13.011038}, \href
  {https://ui.adsabs.harvard.edu/abs/2023PhRvX..13a1038P} {13, 011038}

\bibitem[\protect\citeauthoryear{{Pranav}, {Edelsbrunner}, {van de Weygaert},
  {Vegter}, {Kerber}, {Jones}  \& {Wintraecken}}{{Pranav}
  et~al.}{2017}]{2017MNRAS.465.4281P}
{Pranav} P.,  {Edelsbrunner} H.,  {van de Weygaert} R.,  {Vegter} G.,  {Kerber}
  M.,  {Jones} B. J.~T.,   {Wintraecken} M.,  2017, \mn@doi [\mnras]
  {10.1093/mnras/stw2862}, \href
  {https://ui.adsabs.harvard.edu/abs/2017MNRAS.465.4281P} {465, 4281}

\bibitem[\protect\citeauthoryear{{Pranav} et~al.,}{{Pranav}
  et~al.}{2019a}]{pranav2019topology}
{Pranav} P.,  et~al., 2019a, \mn@doi [\mnras] {10.1093/mnras/stz541}, \href
  {https://ui.adsabs.harvard.edu/abs/2019MNRAS.485.4167P} {485, 4167}

\bibitem[\protect\citeauthoryear{{Pranav}, {Adler}, {Buchert}, {Edelsbrunner},
  {Jones}, {Schwartzman}, {Wagner}  \& {van de Weygaert}}{{Pranav}
  et~al.}{2019b}]{2019A&A...627A.163P}
{Pranav} P.,  {Adler} R.~J.,  {Buchert} T.,  {Edelsbrunner} H.,  {Jones} B.
  J.~T.,  {Schwartzman} A.,  {Wagner} H.,   {van de Weygaert} R.,  2019b,
  \mn@doi [\aap] {10.1051/0004-6361/201834916}, \href
  {https://ui.adsabs.harvard.edu/abs/2019A&A...627A.163P} {627, A163}

\bibitem[\protect\citeauthoryear{{Reed}, {Gardner}, {Quinn}, {Stadel},
  {Fardal}, {Lake}  \& {Governato}}{{Reed} et~al.}{2003}]{2003MNRAS.346..565R}
{Reed} D.,  {Gardner} J.,  {Quinn} T.,  {Stadel} J.,  {Fardal} M.,  {Lake} G.,
   {Governato} F.,  2003, \mn@doi [Monthly Notices of the Royal Astronomical
  Society] {10.1046/j.1365-2966.2003.07113.x}, \href
  {https://ui.adsabs.harvard.edu/abs/2003MNRAS.346..565R} {346, 565}

\bibitem[\protect\citeauthoryear{{Rossi}}{{Rossi}}{2022}]{2022AAS...24031208R}
{Rossi} G.,  2022, in American Astronomical Society Meeting \#240. p. 312.08

\bibitem[\protect\citeauthoryear{Schuster, Hamaus, Pisani, Carbone, Kreisch,
  Pollina  \& Weller}{Schuster et~al.}{2019}]{Schuster2019TheBO}
Schuster N.,  Hamaus N.,  Pisani A.,  Carbone C.,  Kreisch C.~D.,  Pollina G.,
   Weller J.,  2019, Journal of Cosmology and Astroparticle Physics, 2019, 055

\bibitem[\protect\citeauthoryear{{Tanseri}, {Hagstotz}, {Vagnozzi}, {Giusarma}
  \& {Freese}}{{Tanseri} et~al.}{2022}]{2022JHEAp..36....1T}
{Tanseri} I.,  {Hagstotz} S.,  {Vagnozzi} S.,  {Giusarma} E.,   {Freese} K.,
  2022, \mn@doi [Journal of High Energy Astrophysics]
  {10.1016/j.jheap.2022.07.002}, \href
  {https://ui.adsabs.harvard.edu/abs/2022JHEAp..36....1T} {36, 1}

\bibitem[\protect\citeauthoryear{Tejero-Cantero, Boelts, Deistler, Lueckmann,
  Durkan, Goncalves, Greenberg  \& Macke}{Tejero-Cantero
  et~al.}{2020}]{tejero2020sbi}
Tejero-Cantero A.,  Boelts J.,  Deistler M.,  Lueckmann J.,  Durkan C.,
  Goncalves P.,  Greenberg D.,   Macke J.,  2020, The Journal of Open Source
  Software, 5, 2505

\bibitem[\protect\citeauthoryear{Tr\"oster et~al.}{Tr\"oster
  et~al.}{2021}]{KiDS:2020ghu}
Tr\"oster T.,  et~al., 2021, \mn@doi [Astron. Astrophys.]
  {10.1051/0004-6361/202039805}, 649, A88

\bibitem[\protect\citeauthoryear{{Tsizh}, {Tymchyshyn}  \& {Vazza}}{{Tsizh}
  et~al.}{2023}]{2023MNRAS.522.2697T}
{Tsizh} M.,  {Tymchyshyn} V.,   {Vazza} F.,  2023, \mn@doi [Monthly Notices of
  the RAS] {10.1093/mnras/stad1121}, \href
  {https://ui.adsabs.harvard.edu/abs/2023MNRAS.522.2697T} {522, 2697}

\bibitem[\protect\citeauthoryear{{Uhlemann}, {Friedrich},
  {Villaescusa-Navarro}, {Banerjee}  \& {Codis}}{{Uhlemann}
  et~al.}{2020}]{2020MNRAS.495.4006U}
{Uhlemann} C.,  {Friedrich} O.,  {Villaescusa-Navarro} F.,  {Banerjee} A.,
  {Codis} S.,  2020, \mn@doi [Monthly Notices of the RAS]
  {10.1093/mnras/staa1155}, \href
  {https://ui.adsabs.harvard.edu/abs/2020MNRAS.495.4006U} {495, 4006}

\bibitem[\protect\citeauthoryear{{Upadhye}, {Kwan}, {Pope}, {Heitmann},
  {Habib}, {Finkel}  \& {Frontiere}}{{Upadhye}
  et~al.}{2016}]{2016PhRvD..93f3515U}
{Upadhye} A.,  {Kwan} J.,  {Pope} A.,  {Heitmann} K.,  {Habib} S.,  {Finkel}
  H.,   {Frontiere} N.,  2016, \mn@doi [\prd] {10.1103/PhysRevD.93.063515},
  \href {https://ui.adsabs.harvard.edu/abs/2016PhRvD..93f3515U} {93, 063515}

\bibitem[\protect\citeauthoryear{{Vafaei Sadr} \& {Movahed}}{{Vafaei Sadr} \&
  {Movahed}}{2021}]{2021MNRAS.503..815V}
{Vafaei Sadr} A.,  {Movahed} S.~M.~S.,  2021, \mn@doi [Monthly Notices of the
  RAS] {10.1093/mnras/stab368}, \href
  {https://ui.adsabs.harvard.edu/abs/2021MNRAS.503..815V} {503, 815}

\bibitem[\protect\citeauthoryear{{Vagnozzi}, {Giusarma}, {Mena}, {Freese},
  {Gerbino}, {Ho}  \& {Lattanzi}}{{Vagnozzi}
  et~al.}{2017}]{2017PhRvD..96l3503V}
{Vagnozzi} S.,  {Giusarma} E.,  {Mena} O.,  {Freese} K.,  {Gerbino} M.,  {Ho}
  S.,   {Lattanzi} M.,  2017, \mn@doi [Physical Review D]
  {10.1103/PhysRevD.96.123503}, \href
  {https://ui.adsabs.harvard.edu/abs/2017PhRvD..96l3503V} {96, 123503}

\bibitem[\protect\citeauthoryear{{Vagnozzi}, {Dhawan}, {Gerbino}, {Freese},
  {Goobar}  \& {Mena}}{{Vagnozzi} et~al.}{2018a}]{2018PhRvD..98h3501V}
{Vagnozzi} S.,  {Dhawan} S.,  {Gerbino} M.,  {Freese} K.,  {Goobar} A.,
  {Mena} O.,  2018a, \mn@doi [Physical Review D] {10.1103/PhysRevD.98.083501},
  \href {https://ui.adsabs.harvard.edu/abs/2018PhRvD..98h3501V} {98, 083501}

\bibitem[\protect\citeauthoryear{{Vagnozzi}, {Brinckmann}, {Archidiacono},
  {Freese}, {Gerbino}, {Lesgourgues}  \& {Sprenger}}{{Vagnozzi}
  et~al.}{2018b}]{2018JCAP...09..001V}
{Vagnozzi} S.,  {Brinckmann} T.,  {Archidiacono} M.,  {Freese} K.,  {Gerbino}
  M.,  {Lesgourgues} J.,   {Sprenger} T.,  2018b, \mn@doi [Journal of Cosmology
  and Astroparticle Physics] {10.1088/1475-7516/2018/09/001}, \href
  {https://ui.adsabs.harvard.edu/abs/2018JCAP...09..001V} {2018, 001}

\bibitem[\protect\citeauthoryear{{Vielzeuf}, {Calabrese}, {Carbone}, {Fabbian}
  \& {Baccigalupi}}{{Vielzeuf} et~al.}{2023}]{Vielzeuf2023DEMNUniTI}
{Vielzeuf} P.,  {Calabrese} M.,  {Carbone} C.,  {Fabbian} G.,   {Baccigalupi}
  C.,  2023, \mn@doi [Journal of Cosmology and Astroparticle Physics]
  {10.1088/1475-7516/2023/08/010}, \href
  {https://ui.adsabs.harvard.edu/abs/2023JCAP...08..010V} {2023, 010}

\bibitem[\protect\citeauthoryear{{Villaescusa-Navarro}}{{Villaescusa-Navarro}}{2018}]{villaescusa2018pylians}
{Villaescusa-Navarro} F.,  2018, {Pylians: Python libraries for the analysis of
  numerical simulations}, Astrophysics Source Code Library (\mn@eprint {ascl}
  {1811.008})

\bibitem[\protect\citeauthoryear{{Villaescusa-Navarro}, {Marulli}, {Viel},
  {Branchini}, {Castorina}, {Sefusatti}  \& {Saito}}{{Villaescusa-Navarro}
  et~al.}{2014}]{2014JCAP...03..011V}
{Villaescusa-Navarro} F.,  {Marulli} F.,  {Viel} M.,  {Branchini} E.,
  {Castorina} E.,  {Sefusatti} E.,   {Saito} S.,  2014, \mn@doi [Journal of
  Cosmology and Astroparticle Physics] {10.1088/1475-7516/2014/03/011}, \href
  {https://ui.adsabs.harvard.edu/abs/2014JCAP...03..011V} {2014, 011}

\bibitem[\protect\citeauthoryear{{Villaescusa-Navarro}
  et~al.,}{{Villaescusa-Navarro} et~al.}{2020}]{Quijote_sims}
{Villaescusa-Navarro} F.,  et~al., 2020, \mn@doi [Astrophysical Journal s]
  {10.3847/1538-4365/ab9d82}, \href
  {https://ui.adsabs.harvard.edu/abs/2020ApJS..250....2V} {250, 2}

\bibitem[\protect\citeauthoryear{Wasserman}{Wasserman}{2018}]{wasserman2018topological}
Wasserman L.,  2018, Annual Review of Statistics and Its Application, 5, 501

\bibitem[\protect\citeauthoryear{{Wilding}, {Nevenzeel}, {van de Weygaert},
  {Vegter}, {Pranav}, {Jones}, {Efstathiou}  \& {Feldbrugge}}{{Wilding}
  et~al.}{2021}]{2021MNRAS.507.2968W}
{Wilding} G.,  {Nevenzeel} K.,  {van de Weygaert} R.,  {Vegter} G.,  {Pranav}
  P.,  {Jones} B. J.~T.,  {Efstathiou} K.,   {Feldbrugge} J.,  2021, \mn@doi
  [Monthly Notices of the RAS] {10.1093/mnras/stab2326}, \href
  {https://ui.adsabs.harvard.edu/abs/2021MNRAS.507.2968W} {507, 2968}

\bibitem[\protect\citeauthoryear{{Wong}}{{Wong}}{2011}]{2011ARNPS..61...69W}
{Wong} Y. Y.~Y.,  2011, \mn@doi [Annual Review of Nuclear and Particle Science]
  {10.1146/annurev-nucl-102010-130252}, \href
  {https://ui.adsabs.harvard.edu/abs/2011ARNPS..61...69W} {61, 69}

\bibitem[\protect\citeauthoryear{{Xu}, {Cisewski-Kehe}, {Green}  \&
  {Nagai}}{{Xu} et~al.}{2019}]{Xu2019}
{Xu} X.,  {Cisewski-Kehe} J.,  {Green} S.~B.,   {Nagai} D.,  2019, \mn@doi
  [Astronomy and Computing] {10.1016/j.ascom.2019.02.003}, \href
  {https://ui.adsabs.harvard.edu/abs/2019A&C....27...34X} {27, 34}

\bibitem[\protect\citeauthoryear{{Yip}, {Biagetti}, {Cole}, {Viswanathan}  \&
  {Shiu}}{{Yip} et~al.}{2024}]{2024arXiv240313985Y}
{Yip} J. H.~T.,  {Biagetti} M.,  {Cole} A.,  {Viswanathan} K.,   {Shiu} G.,
  2024, \mn@doi [arXiv e-prints] {10.48550/arXiv.2403.13985}, \href
  {https://ui.adsabs.harvard.edu/abs/2024arXiv240313985Y} {p. arXiv:2403.13985}

\bibitem[\protect\citeauthoryear{{Zennaro}, {Bel}, {Villaescusa-Navarro},
  {Carbone}, {Sefusatti}  \& {Guzzo}}{{Zennaro}
  et~al.}{2017}]{2017MNRAS.466.3244Z}
{Zennaro} M.,  {Bel} J.,  {Villaescusa-Navarro} F.,  {Carbone} C.,  {Sefusatti}
  E.,   {Guzzo} L.,  2017, \mn@doi [Monthly Notices of the RAS]
  {10.1093/mnras/stw3340}, \href
  {https://ui.adsabs.harvard.edu/abs/2017MNRAS.466.3244Z} {466, 3244}

\bibitem[\protect\citeauthoryear{{Zhan}}{{Zhan}}{2011}]{2011SSPMA}
{Zhan} H.,  2011, \mn@doi [Scientia Sinica Physica, Mechanica \& Astronomica]
  {10.1360/132011-961}, \href
  {https://ui.adsabs.harvard.edu/abs/2011SSPMA..41.1441Z} {41, 1441}

\bibitem[\protect\citeauthoryear{{Zhang}, {Li}, {Liu}, {Spergel}, {Kreisch},
  {Pisani}  \& {Wandelt}}{{Zhang} et~al.}{2020}]{2020PhRvD.102h3537Z}
{Zhang} G.,  {Li} Z.,  {Liu} J.,  {Spergel} D.~N.,  {Kreisch} C.~D.,  {Pisani}
  A.,   {Wandelt} B.~D.,  2020, \mn@doi [\prd] {10.1103/PhysRevD.102.083537},
  \href {https://ui.adsabs.harvard.edu/abs/2020PhRvD.102h3537Z} {102, 083537}

\bibitem[\protect\citeauthoryear{Zomorodian}{Zomorodian}{2005}]{zomorodian2005topology11}
Zomorodian A.~J.,  2005, Topology for computing.
Cambridge university press

\bibitem[\protect\citeauthoryear{{van de Weygaert} et~al.,}{{van de Weygaert}
  et~al.}{2011}]{van2011}
{van de Weygaert} R.,  et~al., 2011, in , Vol.~6970, Lecture Notes in Computer
  Science.
pp 60--101, \mn@doi{10.1007/978-3-642-25249-5_3}

\makeatother
\end{thebibliography}

\end{document}